\documentclass[aps,prd,nofootinbib,amsmath,amssymb,superscriptaddress,twocolumn]
{revtex4-2}
\usepackage{lineno}
\usepackage{graphics,epsfig,subfigure}
\usepackage{ulem}
\usepackage{booktabs}
\usepackage{color}
\usepackage{url}
\usepackage{orcidlink}
\usepackage{float}
\usepackage{dcolumn}
\usepackage{bm}
\usepackage{latexsym}
\usepackage{booktabs}
\usepackage{amsmath}
\usepackage{amssymb}
\allowdisplaybreaks[4]
\usepackage{multirow}
\newcommand{\be}{\begin{equation}}
\newcommand{\ee}{\end{equation}}
\newcommand{\bq}{\begin{eqnarray}}
\newcommand{\eq}{\end{eqnarray}}
\usepackage{graphicx}
\usepackage{amssymb}
\usepackage{epstopdf}
\usepackage{color}
\usepackage{float}
\usepackage{amsmath}
\usepackage{orcidlink}
\usepackage{latexsym}
\usepackage{amsmath}
\usepackage{amssymb}
\usepackage{amsfonts}
\hypersetup{
     breaklinks=true,
    pdfstartview={FitH},    
    colorlinks=true,       
    linkcolor=blue,          
    citecolor=red,        
    filecolor=magenta,      
    urlcolor=blue,           
    anchorcolor=green,      
    linktocpage=true
}

\usepackage{subfigure}
\usepackage{xcolor}

\begin{document}

\title{Scattering, Hawking Radiation and Neutrino Energy Deposition in Euler-Heisenberg Black Holes Surrounded by Perfect Fluid Dark Matter}

\author{Ram\'on B\'ecar \orcidlink{0000-0003-0590-3957}}
\email{rbecar@uct.cl} \affiliation{\small{Departamento de Ciencias Matem\'aticas y F\'{i}sicas, Facultad de Ingenier\'ia, Universidad Cat\'olica de Temuco, Montt 56, Casilla 15-D, Temuco, Chile.}}

\author{P. A. Gonz\'{a}lez \orcidlink{0000-0002-9685-4021}}
\email{pablo.gonzalez@udp.cl} \affiliation{Facultad de
Ingenier\'{i}a y Ciencias, Universidad Diego Portales, Avenida Ej\'{e}rcito
Libertador 441, Casilla 298-V, Santiago, Chile.}

\author{Ali \"Ovg\"un \orcidlink{0000-0002-9889-342X}}
\email{ali.ovgun@emu.edu.tr}
\affiliation{Physics Department, Faculty of Arts and Sciences, Eastern Mediterranean University, Famagusta, 99628 North Cyprus via Mersin 10, Turkiye.}

\author{Joel Saavedra \orcidlink{0000-0002-1430-3008}}
\email{joel.saavedra@pucv.cl }
\affiliation{
Instituto de F\'isica,
Pontificia Universidad Cat\'olica de Valpara\'iso,
Casilla 4059, Valpara\'iso, Chile
}

\author{Yerko V\'asquez \orcidlink{0000-0003-1762-6678}}
\email{yvasquez@userena.cl}
\affiliation{Departamento de F\'isica, Facultad de Ciencias, Universidad de La Serena,\\
Avenida Cisternas 1200, La Serena, Chile.}

\date{\today}

\begin{abstract}
We study the dynamical and scattering properties for the Euler-Heisenberg black holes surrounded by perfect fluid dark matter. The geometry contains a compact non-linear electrodynamic correction governed by the Euler-Heisenberg coupling and a logarithmic dark-matter contribution governed by the surrounding PFDM halo. We study the scalar, electromagnetic and a particular effective axial spin-2 channel constructed on the fixed Euler-Heisenberg plus PFDM background, acting as a proxy for the gravitational-like perturbation problem and not for the fully coupled gravitational perturbation problem. We compute the quasinormal-mode spectrum employing a thirteenth-order WKB method supplemented with Pad\'e resummation and compare it with the eikonal prediction calculated in terms of the angular frequency of the photon sphere and the Lyapunov exponent. Moreover, we study the near-extremal configurations and derive a purely imaginary branch of quasinormal frequencies in the near-horizon region, whose damping rate increases with the PFDM parameter and is nearly spin-independent. We then compute exact greybody factors by direct numerical integration of the radial wave equation and compare them to analytical lower bounds. We also analyze the absorption cross sections and the Hawking emission spectra. We also calculate the relativistic enhancement of the neutrino-antineutrino annihilation channel $\nu\bar\nu\rightarrow e^{-}e^{+}$ outside the Euler-Heisenberg-PFDM black hole.  We find that the PFDM parameter contracts the optical structure, increases the oscillation frequency, enhances the damping rate and suppresses transmission. On the other hand, the Euler-Heisenberg correction leads to a weaker near-horizon deformation whose effect becomes relevant for sufficiently large charge. These results provide a common scattering framework for comparing the impact of dark-matter environments and nonlinear electrodynamics on black-hole spectroscopic and radiative observables.
\end{abstract}

\maketitle

\tableofcontents

\clearpage

\section{Introduction}

The theory of black hole perturbations provides a straightforward way to probe the strong-field regime of gravity. The late-time response of a compact object in the aftermath of a perturbation is characterized by a discrete set of damped oscillations, the quasinormal modes (QNMs), whose complex frequencies depend on the background geometry, the spin of the perturbing field and the boundary conditions applied at the event horizon and at infinity \cite{Berti:2009kk,Kokkotas:1999bd,Konoplya:2011qq}. Also the wave scattering, absorption, and greybody factors \cite{Page:1976df,Boonserm:2008zg,Boonserm:2013sza} are governed by the same effective potentials that govern the ringdown phase. These greybody factors modify the pure thermal Hawking spectrum at infinity \cite{Hawking:1975vcx,Page:1976df}. Thus ringdown spectra, greybody factors, absorption cross sections, Hawking radiation and neutrino energy deposition are complementary probes of black-hole spacetimes and their optical, dynamical and radiative properties.

The eikonal correspondence gives a helpful connection between wave behaviour and geometrical optics. In the large angular-momentum limit, the real part of the QNM frequency is set by the angular velocity of the unstable circular null orbit, while the imaginary part is fixed by the corresponding Lyapunov exponent \cite{Cardoso:2008bp}. The photon sphere is therefore at the heart of shadows and lensing, but also of ringdown and high-frequency absorption and scattering phenomena. This link is especially useful in modified black-hole geometries where environmental effects or nonlinear field corrections may imprint correlated signatures in multiple observables.

Astrophysical black holes are not expected to be isolated; they may be embedded in accretion flows, plasma environments or dark matter halos. Black holes with perfect fluid dark matter (PFDM) offer a phenomenologically tractable description of dark matter effects with a logarithmic radial correction to the lapse function that modifies the horizon structure, photon sphere, effective potentials, and thermodynamic properties \cite{Ma:2024oqe,Becar:2023jtd}. In the previous work \cite{Becar:2023jtd} we have demonstrated the effects of PFDM on quasinormal spectra and optical observables which motivates a systematic study of the effects of PFDM on scattering and radiative processes.

Another such deviation from the standard Reissner-Nordstr\"om geometry is due to the nonlinear electrodynamics.
The leading quantum-electrodynamic corrections due to vacuum polarization in strong electromagnetic fields are included in the Euler-Heisenberg effective theory \cite{Heisenberg:1936nmg,Schwinger:1951nm}. When these corrections are coupled to gravity, they deform the charged black-hole geometry by short-range terms that are relevant in the strong-field region \cite{Breton:2021zsd}. Unlike the long-range PFDM contribution, the Euler-Heisenberg correction is compact and decays rapidly with radius. Thus the combined Euler–Heisenberg–PFDM geometry serves as a useful theoretical laboratory for disentangling near-horizon nonlinear-electrodynamic effects from extended dark matter environmental effects.

The Euler–Heisenberg black hole in the PFDM background was first constructed in Ref.~\cite{Ma:2024oqe} and its optical and thermodynamic properties were studied. The photon-sphere structure, shadow, eikonal QNMs, sparsity and energy emission rate were investigated later in Ref.~\cite{Silva:2026}. Ref.~\cite{Belchior:2026} explored scalar, electromagnetic, and Dirac perturbations and Boonserm-Visser lower bounds for the greybody factors and partial absorption cross sections. In a more recent work, scalar and electromagnetic QNMs were computed by the asymptotic iteration method and sixth-order WKB approximation in Ref. \cite{Feng:2026}. These studies demonstrate the modification of the effective potential barriers and the corresponding transmission probabilities by both the PFDM and Euler-Heisenberg parameters. But a consistent description of the photon-sphere properties, the quasinormal ringdown, the exact scattering, the Hawking radiation and the neutrino energy deposition in the same Euler-Heisenberg-PFDM geometry is still missing.

The goal of the present work is to provide such a unified framework. We investigate scalar, electromagnetic, and effective axial spin-2 perturbations of the Euler-Heisenberg black hole in the presence of PFDM. The scalar and electromagnetic sectors are considered test field perturbations of the fixed background. The spin-2 sector is governed by an effective Regge-Wheeler-like axial potential. This choice is inspired by the standard treatment of the Schwarzschild effective axial spin-2 perturbations \cite{Regge:1957td,Chandrasekhar:1983}. We emphasize that this effective spin-2 channel is a phenomenological surrogate for gravitational-like perturbations on the fixed EH+PFDM geometry, not the outcome of a fully coupled gravitational-electromagnetic-dark-matter perturbation analysis.

We calculate quasinormal frequencies via the thirteenth-order Pad\'e-resummed WKB approximation \cite{Matyjasek:2017psv,Konoplya:2019hlu} and compare with the eikonal prediction from the photon-sphere angular frequency and Lyapunov exponent \cite{Cardoso:2008bp}. We also analyze the near-extremal regime where we investigate the emergence of purely imaginary quasinormal frequencies induced by the near-horizon geometry.

The exact greybody factors are obtained by direct numerical integration of the radial wave equation. The transmission coefficient is obtained from the asymptotic scattering solution. These transmission probabilities are then used to compute the absorption cross sections and the Hawking emission spectra to quantify the effect of PFDM and nonlinear electrodynamic corrections on the reflection-transmission transition, the high-frequency absorption regime and the thermal radiation profile.

Finally, we discuss the relativistic enhancement of neutrino-antineutrino annihilation ($\nu+\bar{\nu}\rightarrow e^-+e^+$), a mechanism that has long been discussed in the context of compact objects and gamma-ray-burst central engines \cite{Salmonson:1999es,Asano:2000dq,Asano:2000ib}. The local rate of deposition is sensitive to the redshifted neutrino temperature, and thereby provides a useful probe of the near horizon geometry.
We determine the corresponding Newtonian-normalized deposition rate and explore the relative importance of the electric charge, the Euler-Heisenberg coupling and the PFDM parameter. This provides a direct connection between environmental dark-matter effects, nonlinear electrodynamics, wave scattering, thermal emission, and relativistic energy-deposition processes within a single black-hole framework.

The paper is structured as follows. In Sec.~\ref{EH} we consider the Euler-Heisenberg (EH) black hole surrounded by PFDM and analyze the horizon structure. In Sec. \ref{EffPot} we derive the effective potentials and discuss the photon-sphere quantities. In Sec. \ref{QNMs} we compute the quasinormal spectrum and discuss the eikonal and near-extremal limits. In Sec.~\ref{GB} we calculate exact greybody factors numerically. Sec.~\ref{ACS} shows the absorption cross sections and the Hawking emission spectra. In Sec.~\ref{sec:nu_nubar_eh_pfdm} we discuss the energy deposition rate from neutrino-antineutrino annihilation. Finally, we summarize our conclusions in Sec.~\ref{conclusions}.

\section{Euler-Heisenberg Black Hole Surrounded by Perfect Fluid Dark Matter}
\label{EH}
We consider a static and spherically symmetric black-hole spacetime
described by
\begin{equation}
ds^2
=
-f(r)dt^2
+
\frac{dr^2}{f(r)}
+
r^2 d\Omega^2 .
\label{eq:metric}
\end{equation}
The metric function is \cite{Ma:2024oqe}
\begin{equation}
f(r)
=
1
-
\frac{2M}{r}
+
\frac{Q^2}{r^2}
-
\frac{aQ^4}{20r^6}
+
\frac{\alpha}{r}
\ln\left(\frac{r}{|\alpha|}\right),
\label{eq:fmetric}
\end{equation}
where $M$ is the mass, $Q$ is the electric charge, $a$ is the
Euler-Heisenberg coupling, and $\alpha$ characterizes the PFDM halo.

The term $Q^2/r^2$ is the standard Reissner-Nordstr\"om contribution.
The term
\begin{equation}
-\frac{aQ^4}{20r^6}
\end{equation}
is the leading Euler-Heisenberg correction. It is short-ranged and
therefore mainly affects the near-horizon and photon-sphere regions.
The PFDM term,
\begin{equation}
\frac{\alpha}{r}\ln\left(\frac{r}{|\alpha|}\right),
\end{equation}
has a logarithmic structure and modifies the geometry on larger scales.

\begin{figure}[h]
\centering
\includegraphics[width=\columnwidth]{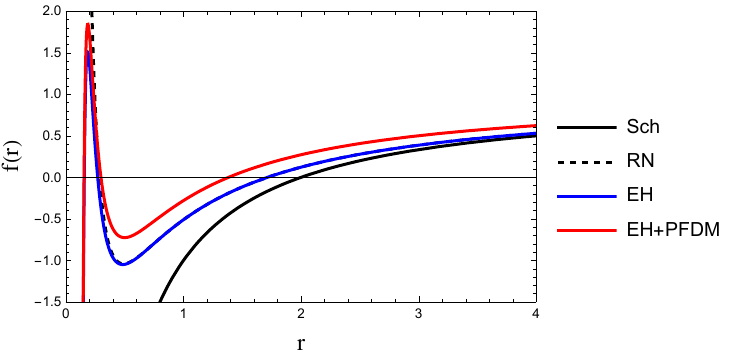}
\caption{
Metric function $f(r)$ as a function of the radial coordinate $r$ for the Schwarzschild (Sch), Reissner-Nordstr\"om (RN) with $Q=0.7$, Euler-Heisenberg (EH) with $Q=0.7$, and $a=0.01$, and Euler-Heisenberg black hole surrounded by perfect fluid dark matter (EH+PFDM) with $Q=0.7$, $a=0.01$, and $\alpha=0.1$. 
}
\label{LF}
\end{figure}

The event horizon radius $r_h$ is given by the largest positive root of
\begin{equation}
f(r_h)=0 .
\label{eq:horizon}
\end{equation}
The geometry reduces to Reissner-Nordstr\"om for $a=\alpha=0$, to Schwarzschild for $Q=a=\alpha=0$ and to the pure Euler-Heisenberg black hole for $\alpha=0$ and $a\neq 0$.
In this work we use units in which $M=1$ so that the parameter $\alpha$ is a dimensionless measure of the PFDM deformation. We employ the representative range $0 \leq \alpha \leq 0.3$ for exploratory purposes to show how the optical, dynamical and thermodynamic properties of the black hole are modified by the presence of an ambient dark matter halo. The purpose of the analysis is hence to emphasize qualitative tendencies rather than directly fit astrophysical dark matter distributions. Fig. \ref{LF} shows the metric function $f(r)$ for the Schwarzschild, Reissner-Nordstr\"om (RN), Euler-Heisenberg (EH) and EH black holes with perfect fluid dark matter (EH+PFDM). 
As expected the Schwarzschild solution has a single event horizon at $r_h=2M$. The RN geometry on the other hand has the usual two-horizon structure of the charged black hole with an inner Cauchy horizon and an outer event horizon. The nonlinear corrections of the Euler-Heisenberg type lead to drastic changes in the causal structure of the spacetime. The EH solution shows three positive real roots for the parameters considered, indicating the existence of an additional inner horizon due to the nonlinear electromagnetic sector. The outer horizon is close to the RN horizon but the additional roots appear in the strong field regime where the term $Q^4/r^6$ dominates. The multihorizon structure is preserved in the EH+PFDM geometry, but the location of the inner horizons is shifted by the logarithmic PFDM term. These results suggest that the effects of nonlinear electrodynamics and dark matter can significantly enrich the horizon structure as compared to the standard RN case, leading to geometries with multiple internal causal boundaries.

\begin{figure}[h]
\centering
\includegraphics[width=0.47\textwidth]{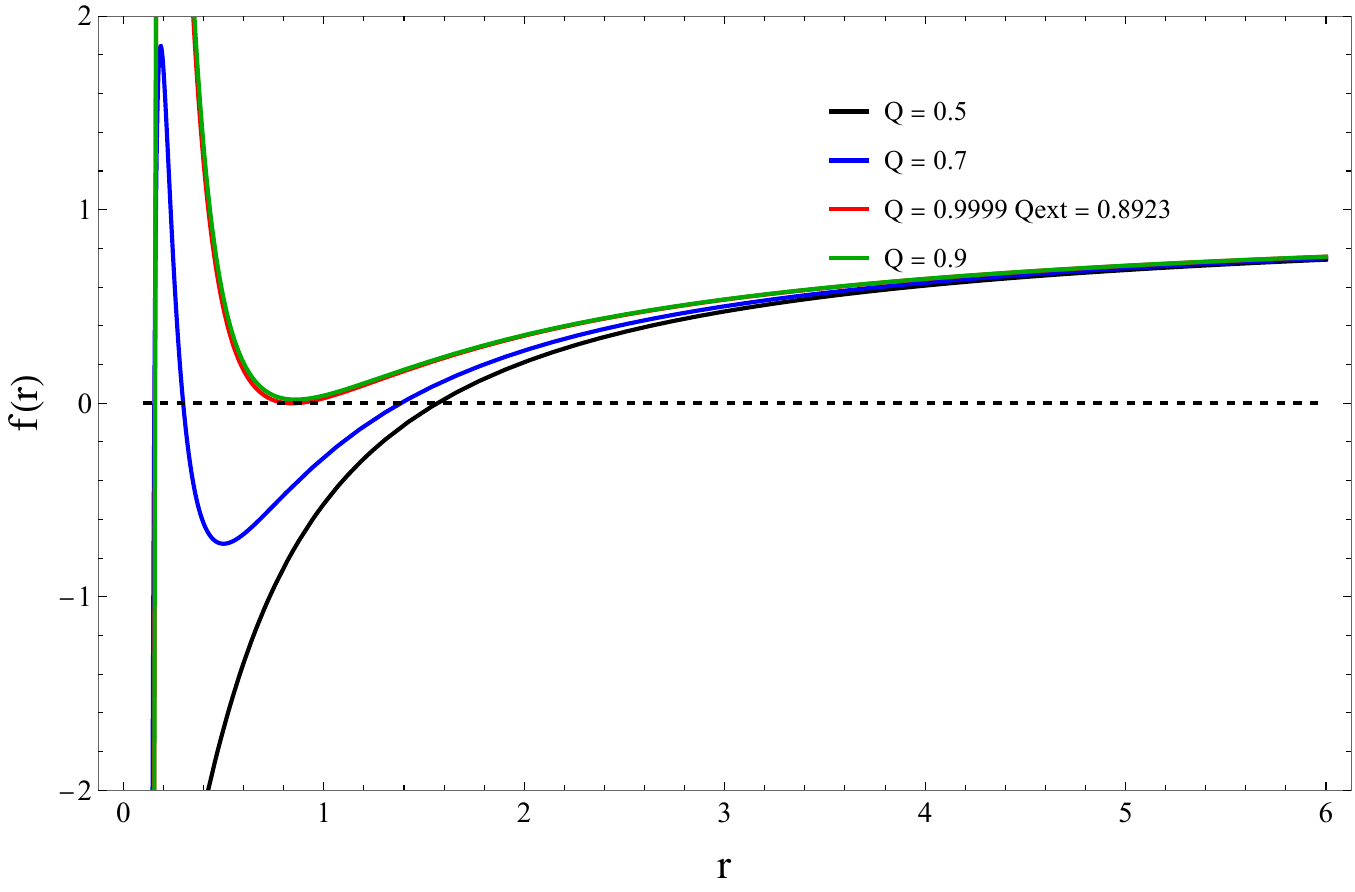}
\caption{
Metric function $f(r)$ of the EH+PFDM black hole for $Q=0.5,0.7,0.9,0.9999Q_{ext}$, with $M=1$, $a=0.01$, and $\alpha=0.1$. The zeros of $f(r)$ indicate the horizons. For sufficiently large charge, the inner and outer horizons approach each other, indicating a near-extremal configuration.
}
\label{LF2}
\end{figure}

Fig. \ref{LF2} further illustrates how the horizon structure of the
EH+PFDM geometry changes as the electric charge approaches its
extremal value. In this plot we keep \(M=1\), \(a=0.01\), and
\(\alpha=0.1\) fixed, and vary \(Q\). For moderate charges, the
metric function displays the characteristic multihorizon structure
already discussed above. As \(Q\) is increased, the inner and outer
horizons move closer to each other, and the minimum of \(f(r)\) near
the horizon approaches zero. The curve
\(Q=0.9999Q_{\rm ext}\) represents a near-extremal configuration, in
which the two relevant horizons are almost degenerate. This behavior
provides the geometrical motivation for the near-extremal analysis
performed in Sec.~\ref{NEM}, where the small surface gravity of the nearly
degenerate horizon gives rise to a branch of purely imaginary
quasinormal frequencies. Thus, Fig. \ref{LF2} connects the horizon structure
of the EH+PFDM black hole with the emergence of the near-horizon
modes studied later in the paper.

\section{Effective Potentials and Photon-Sphere Structure}
\label{EffPot}
\begin{figure}[h]
\centering
\includegraphics[width=0.4\textwidth]{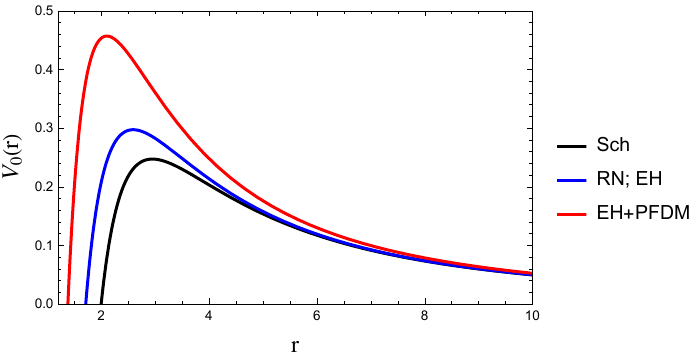}
\includegraphics[width=0.4\textwidth]{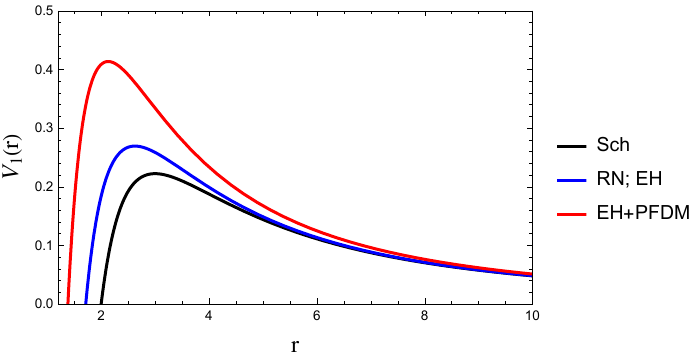}
\includegraphics[width=0.4\textwidth]{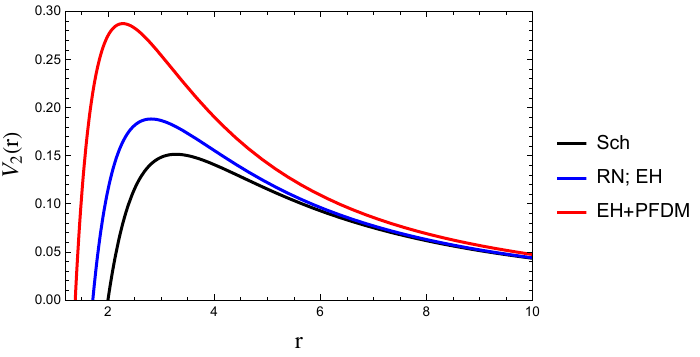}
\caption{
Effective potentials for scalar (top panel), electromagnetic (central panel), and effective
spin-2 perturbations (bottom panel) of the Schwarzschild (Sch), Reissner-Nordstr\"om (RN) with $Q=0.7$, Euler-Heisenberg (EH) with $Q=0.7$, and $a=0.01$, and Euler-Heisenberg black hole surrounded by perfect fluid dark matter (EH+PFDM) with $Q=0.7$, $a=0.01$, and $\alpha=0.1$. Here, $\ell=2$. The RN and EH curves overlap almost completely and are therefore shown as a single curve.
}
\label{Vs}
\end{figure}

Perturbations of the background can be written in Schr\"odinger-like
form as
\begin{equation}
\frac{d^2\Psi}{dr_*^2}
+
\left[
\omega^2
-
V_s(r)
\right]\Psi
=
0,
\label{eq:master}
\end{equation}
where the tortoise coordinate is defined by
\begin{equation}
\frac{dr_*}{dr}
=
\frac{1}{f(r)}.
\end{equation}
For the bosonic sectors considered here, we use the compact
spin-dependent effective potential
\begin{equation}
V_s(r)
=
f(r)
\left[
\frac{\ell(\ell+1)}{r^2}
+
(1-s^2)\frac{f'(r)}{r}
\right].
\label{eq:Vspin}
\end{equation}
For $s=0$ this gives the scalar potential
\begin{equation}
V_0(r)
=
f(r)
\left[
\frac{\ell(\ell+1)}{r^2}
+
\frac{f'(r)}{r}
\right],
\end{equation}
while for $s=1$ it gives the electromagnetic potential
\begin{equation}
V_1(r)
=
f(r)
\frac{\ell(\ell+1)}{r^2}.
\end{equation}
For the spin-2 sector, we adopt an effective axial Regge Wheeler type
potential constructed by analogy with the Schwarzschild case,
\begin{equation}
V_2(r)=
f(r)\left[
\frac{\ell(\ell+1)}{r^2}
-
\frac{3f'(r)}{r}
\right].
\label{V2}
\end{equation}
This expression reduces in the Schwarzschild limit exactly to the usual Regge-Wheeler potential. In the current Euler Heisenberg plus PFDM geometry, however, it should be considered a phenomenological ansatz on the fixed background, rather than the result of a complete coupled gravitational perturbation analysis. It aims at providing an efficient description of gravitational-like perturbations and studying the effect of the modified metric on ringdown and scattering observables. The systematic uncertainty for the spin-2 results is larger than for the scalar and electromagnetic sectors. 
In Fig. \ref{Vs} we show the effective potentials for scalar ($s=0$), electromagnetic ($s=1$) and effective spin-2 perturbations of the Schwarzschild, RN, EH and EH+PFDM. The potentials exhibit in all cases the typical shape of a barrier, necessary for quasinormal oscillations and scattering processes. The height of the potential peak is very sensitive to the nonlinear electromagnetic properties and to the contribution of the PFDM.
For the parameter choices considered here, the Schwarzschild geometry yields the lowest barrier, while the RN case and the EH case give higher peaks due to the presence of the electric charge. The RN and EH curves are almost indistinguishable for the parameters considered, indicating that the nonlinear corrections of the Euler-Heisenberg theory induce small deviations of the effective potential outside the strong-field regime. In the opposite case, the EH+PFDM spacetime results in a significantly higher and thinner barrier in all the spin sectors. This behavior arises because the logarithmic contribution of PFDM into the metric function enters into both the redshift factor and the radial gradient appearing in the effective potentials.
The rise of the potential barrier height has important consequences for the perturbative dynamics. In particular, we find that larger barriers correspond to higher oscillation frequencies, modified damping rates, and stronger backscattering. The effect is especially significant in the scalar and electromagnetic sectors, where the EH+PFDM potential exhibits a considerable upward shift around the photon-sphere region. The spin-2 potential used in this work is phenomenological, but it shares the same qualitative behavior. Hence, the combined effects of nonlinear electrodynamics and the dark matter environment can drastically modify the ringdown structure with respect to the Schwarzschild and RN geometries.
\section{Quasinormal Modes and Eikonal Correspondence}
\label{QNMs}

\subsection{Photon sphere modes}

Before presenting the quasinormal mode analysis, we emphasize that the
spin-2 sector considered in this work is modeled through an effective
axial Regge Wheeler type potential defined on the fixed background
geometry. This construction provides a phenomenological description of
gravitational-like perturbations and is intended to capture the impact
of the modified metric on the ringdown spectrum. It does not represent
the full set of coupled gravitational perturbation equations of the
Euler Heisenberg plus PFDM spacetime.
The QNM boundary conditions are purely ingoing waves at the event
horizon and purely outgoing waves at spatial infinity:
\begin{equation}
\Psi
\sim
e^{-i\omega r_*},
\qquad
r_* \to -\infty,
\end{equation}
and
\begin{equation}
\Psi
\sim
e^{+i\omega r_*},
\qquad
r_* \to +\infty .
\end{equation}

We compute the quasinormal frequencies using the high-order WKB method
supplemented by Pad\'e resummation. The WKB quantization condition at
order $N$ is
\begin{equation}
\frac{i\left(\omega^2-V_0\right)}
{\sqrt{-2V_0''}}
-
\sum_{j=2}^{N}\Lambda_j
=
n+\frac12 ,
\label{eq:WKB}
\end{equation}
where $V_0$ is the maximum of the effective potential, $V_0''$ is its
second derivative with respect to the tortoise coordinate evaluated at
the maximum, and $\Lambda_j$ are higher-order correction terms
depending on derivatives of the potential up to order $2j$.

In this work we take $N=13$ and apply Pad\'e resummation to improve the
convergence of the WKB series. The Pad\'e-WKB13 frequencies are used
as the main numerical QNM data. The method is most reliable for modes
with $\ell>n$, and we therefore focus primarily on fundamental modes
$n=0$ with $\ell\geq2$.

In the eikonal regime, the QNM spectrum is connected with the unstable
photon orbit according to \cite{Cardoso:2008bp}
\begin{equation}
\omega_{\ell n}
\simeq
\ell\Omega_c
-
i\left(n+\frac12\right)\lambda_c .
\label{eq:eikonal}
\end{equation}

The photon sphere is determined by the unstable circular null orbit,
whose radius $r_c$ satisfies
\begin{equation}
2f(r_c)-r_c f'(r_c)=0 .
\label{eq:photonSphere}
\end{equation}
The corresponding angular frequency is
\begin{equation}
\Omega_c
=
\sqrt{\frac{f(r_c)}{r_c^2}},
\label{eq:OmegaC}
\end{equation}
and the Lyapunov exponent is
\begin{equation}
\lambda_c
=
\sqrt{
\frac{
f(r_c)
\left[
2f(r_c)-r_c^2 f''(r_c)
\right]
}{
2r_c^2
}
}.
\label{eq:LambdaC}
\end{equation}

The relation (\ref{eq:eikonal}) connects the oscillation frequency to the angular
frequency of the photon sphere and the damping rate to the instability
timescale of the null orbit.
The PFDM parameter $\alpha$ modifies the photon sphere through the
logarithmic term in Eq.~(\ref{eq:fmetric}). As $\alpha$ increases, the
event horizon and photon sphere are shifted inward, while $\Omega_c$
and $\lambda_c$ increase. Consequently, both the real part and the
magnitude of the imaginary part of the QNM frequencies increase. The
Euler-Heisenberg parameter $a$, by contrast, introduces a more
localized near-horizon correction, whose effect becomes more visible at
larger $Q$.
Table~\ref{tab:QNM_PadeWKB13} shows that increasing the PFDM parameter
systematically decreases $r_h$, $r_c$, and $r_0$ in all three spin
sectors. At the same time, both $\mathrm{Re}(\omega)$ and
$-\mathrm{Im}(\omega)$ increase monotonically. This confirms that the
PFDM halo makes the optical structure more compact and produces a
faster and more strongly damped ringdown signal.

The comparison with the eikonal quantities shows that
$(n+1/2)\lambda_c$ tracks the damping rate with high accuracy,
especially for the scalar and electromagnetic sectors. The real part
follows the same trend as $\ell\Omega_c$, although finite-$\ell$
deviations remain visible for $\ell=2$. These deviations are expected,
since the eikonal correspondence is strictly valid only in the
large-$\ell$ limit.
The deviations from the eikonal prediction are systematically larger in
the effective spin-2 sector than in the scalar and electromagnetic
cases. This behavior has two complementary origins. First, the
correspondence with Eq. (\ref{eq:eikonal})
is formally valid only in the large-$\ell$ limit, whereas the present
analysis focuses on the fundamental mode with 
$(\ell=2)$. Second, the
spin-2 channel is described by an effective axial Regge Wheeler type
potential introduced as a phenomenological ansatz on the fixed
background rather than as the result of a fully coupled gravitational
perturbation analysis. The larger Pad\'e uncertainties and the more
pronounced deviations from the eikonal estimate should therefore be
interpreted as a combination of finite-$\ell$ corrections and the
additional systematic uncertainty associated with the effective
gravitational sector.
\begin{table*}[!t]
\centering
\caption{Fundamental quasinormal frequencies of the Euler-Heisenberg black hole surrounded by perfect fluid dark matter (PFDM), computed with the Pad\'e-WKB method of order 13 for scalar ($s=0$), electromagnetic ($s=1$), and effective axial gravitational ($s=2$) perturbations with $\ell=2$, $n=0$, $Q=0.7$ and $a=0.01$. The table also lists the event horizon radius $r_h$, the photon sphere radius $r_c$, the position $r_0$ of the maximum of the effective potential, and the eikonal quantities $\ell\Omega_c$ and $(n+\tfrac12)\lambda_c$. The last column shows the estimated Pad\'e error.}
\label{tab:QNM_PadeWKB13}
\scriptsize
\begin{tabular}{cccccccccccc}
\hline\hline
$s$ & $\ell$ & $n$ & $\alpha$ & $r_h$ & $r_{c}$ & $r_{0}$ & ${Re}(\omega)$ & $-{Im} (\omega)$ & $\ell\Omega_c$ & $(n+\frac{1}{2})\lambda_c$ & Pad\'e error \\
\hline
0 & 2 & 0 & 0.001 & 1.705213 & 2.613763 & 2.580894 & 0.535155 & 0.098997 & 0.425728 & 0.098536 & 2.261 $\times$ $10^{-10}$ \\
0 & 2 & 0 & 0.050 & 1.504030 & 2.313876 & 2.285564 & 0.605132 & 0.111406 & 0.481274 & 0.110888 & 2.143 $\times 10^{-10}$ \\
0 & 2 & 0 & 0.100 & 1.383046 & 2.131300 & 2.105259 & 0.660313 & 0.121919 & 0.525013 & 0.121341 & 2.666 $\times 10^{-10}$ \\
0 & 2 & 0 & 0.200 & 1.239964 & 1.909963 & 1.885588 & 0.749498 & 0.141061 & 0.595553 & 0.140341 & 6.813 $\times 10^{-10}$\\
0 & 2 & 0 & 0.300 & 1.173290 & 1.799250 & 1.774194 & 0.814386 & 0.158266 & 0.646688 & 0.157367 & 8.977 $\times 10^{-10}$ \\
\hline
1 & 2 & 0 & 0.001 & 1.705213 & 2.613763 & 2.613763 & 0.508459 & 0.097440 & 0.425728 & 0.098536 & 2.983 $\times 10^{-10}$ \\
1 & 2 & 0 & 0.050 & 1.504030 & 2.313876 & 2.313876 & 0.574925 & 0.109639 & 0.481274 & 0.110888 & 1.856 $\times 10^{-10}$ \\
1 & 2 & 0 & 0.100 & 1.383046 & 2.131300 & 2.131300 & 0.627058 & 0.119940 & 0.525013 & 0.121341 & 2.987 $\times 10^{-10}$ \\
1 & 2 & 0 & 0.200 & 1.239964 & 1.909963 & 1.909963 & 0.710538 & 0.138616 & 0.595553 & 0.140341 & 8.145 $\times 10^{-10}$\\
1 & 2 & 0 & 0.300 & 1.173290 & 1.799250 & 1.799250 & 0.770124 & 0.155295 & 0.646688 & 0.157367 & 1.341 $\times 10^{-10}$\\
\hline
2 & 2 & 0 & 0.001 & 1.705213 & 2.613763 & 2.799124 & 0.421721 & 0.091355 & 0.425728 & 0.098536 & 1.589 $\times 10^{-10}$ \\
2 & 2 & 0 & 0.050 & 1.504030 & 2.313876 & 2.474708 & 0.476648 & 0.102606 & 0.481274 & 0.110888 & 7.336 $\times 10^{-10}$ \\
2 & 2 & 0 & 0.100 & 1.383046 & 2.131300 & 2.280996 & 0.518773 & 0.111993 & 0.525013 & 0.121341 & 1.331 $\times 10^{-10}$ \\
2 & 2 & 0 & 0.200 & 1.239964 & 1.909963 & 2.054615 & 0.583571 & 0.128767 & 0.595553 & 0.140341 & 1.835 $\times 10^{-10}$ \\
2 & 2 & 0 & 0.300 & 1.173290 & 1.799250 & 1.953746 & 0.625934 & 0.143533 & 0.646688 & 0.157367 & 1.380 $\times 10^{-10}$ \\
\hline\hline

\end{tabular}
\end{table*}

\begin{figure}[h]
\centering
\includegraphics[width=0.4\textwidth]{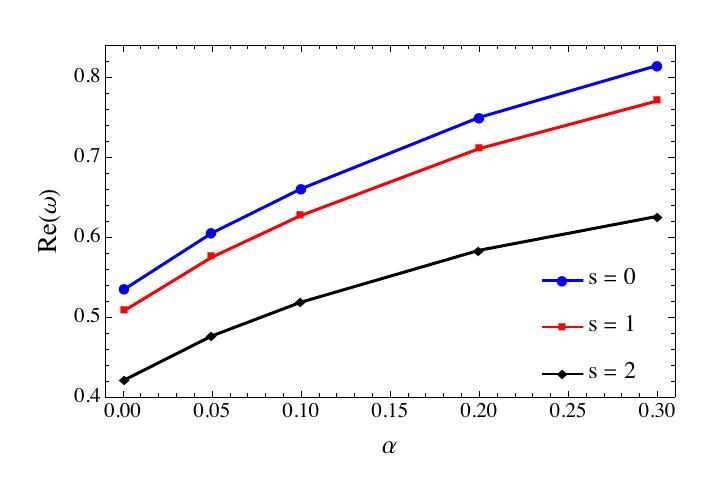}\\
\includegraphics[width=0.4\textwidth]{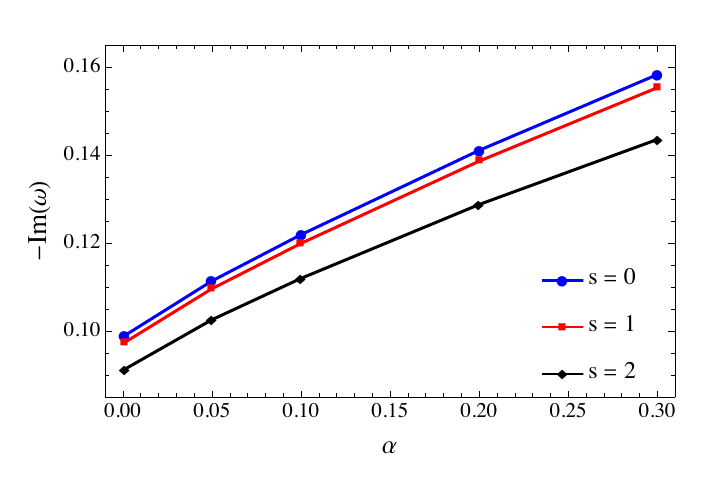}
\caption{
Dependence of the fundamental quasinormal frequencies on the perfect
fluid dark matter parameter $\alpha$ for scalar ($s=0$),
electromagnetic ($s=1$), and effective axial spin-2 ($s=2$)
perturbations with $Q=0.7$, $a=0.01$, $\ell=2$ and $n=0$. Top panel shows the real part
of the frequency, while bottom panel displays the damping rate
$-\mathrm{Im}(\omega)$. }
\label{fig:QNM_vs_alpha}
\end{figure}

\begin{figure}[t]
\centering
\includegraphics[width=0.4\textwidth]{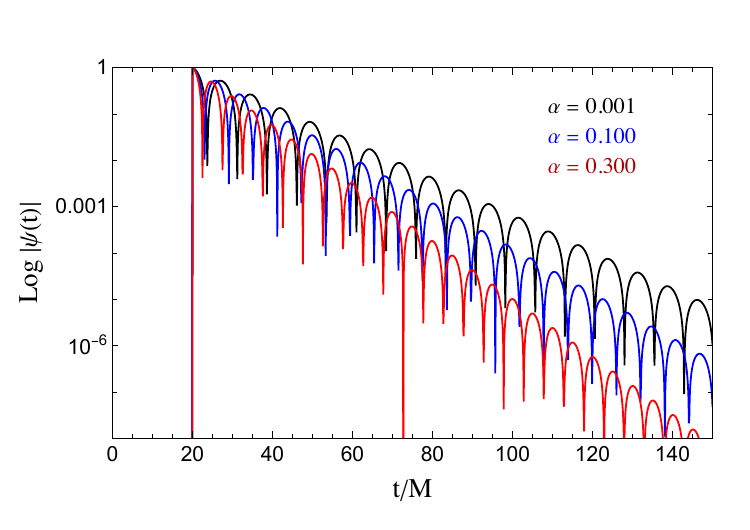}\\
\includegraphics[width=0.4\textwidth]{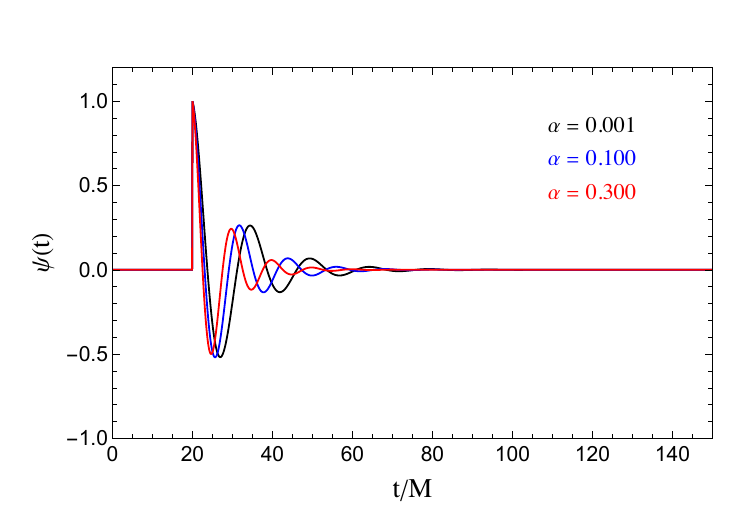}
\caption{
Time domain ringdown waveforms reconstructed from the fundamental
Pad\'e WKB13 quasinormal frequencies of the effective axial spin-2
sector with $Q=0.7$, $a=0.01$, $\ell=2$ and $n=0$ for three representative values of the
perfect fluid dark matter parameter $\alpha$.
Top panel shows the logarithm of the waveform envelope,
$\log|\psi(t)|$, while bottom panel displays the oscillatory signal
$\psi(t)$.
Increasing $\alpha$ leads to both a higher oscillation frequency and a
faster exponential decay, reflecting the simultaneous growth of
$\mathrm{Re}(\omega)$ and $-\mathrm{Im}(\omega)$.
The PFDM halo therefore produces a more compact optical structure and a
shorter-lived ringdown signal.
}
\label{fig:RingdownWaveforms}
\end{figure}

Fig.~\ref{fig:QNM_vs_alpha} complements the numerical results presented
in Table~\ref{tab:QNM_PadeWKB13} by showing the dependence of the
fundamental quasinormal frequencies on the PFDM parameter $\alpha$ for
the scalar ($s=0$), electromagnetic ($s=1$), and effective axial
spin-2 ($s=2$) sectors. The left panel displays the real part of the
frequency, while the right panel shows the damping rate
$-\mathrm{Im}(\omega)$.
This behavior reflects the inward shift
of the event horizon and photon sphere induced by the logarithmic PFDM
term, which increases the photon-sphere angular frequency $\Omega_c$
and the associated Lyapunov exponent $\lambda_c$. 
The scalar sector exhibits the largest frequencies and damping rates,
followed by the electromagnetic sector, whereas the effective
gravitational channel shows systematically smaller values. Despite these
quantitative differences, all sectors display the same qualitative
response to the dark-matter environment.
Fig.~\ref{fig:RingdownWaveforms} provides a time domain
representation of the ringdown signal reconstructed from the
fundamental Pad\'e WKB13 frequencies of the effective axial spin-2
sector. As the PFDM parameter $\alpha$ increases, the oscillation
period decreases and the waveform decays more rapidly, in direct
agreement with the monotonic growth of both $\mathrm{Re}(\omega)$ and
$-\mathrm{Im}(\omega)$ reported in Table~\ref{tab:QNM_PadeWKB13}.
This visualization highlights the observable impact of the dark matter
halo on the temporal structure of the ringdown.

\subsection{Near-extremal modes}
\label{NEM}
Near-extremal modes
emerge when $\epsilon_Q\equiv 1-\frac{Q}{Q_{\rm ext}}\ll 1$. This regime is tightly connected to previous work on perturbations of near-extremal black holes where different families of QNMs were found \cite{Cardoso:2017soq,Fontana:2020syy,Gonzalez:2022upu,Becar:2024agj}. Since the lowest-lying frequency is given directly by the gravitational surface of the event horizon, the purely imaginary solutions will be small in comparison to the PS solution (consequently dominant) whenever the temperature of the black hole is small.
We now turn to the near-extremal branch of purely imaginary
quasinormal frequencies. For each value of the PFDM parameter
\(\alpha\), the extremal configuration is obtained by solving
\[
f(r_{\rm ext},Q_{\rm ext})=0,
\qquad
\partial_r f(r_{\rm ext},Q_{\rm ext})=0 .
\]
The near-extremal charge used in the numerical calculation is then
chosen as
\[
Q_{\rm near}=0.9999\,Q_{\rm ext},
\qquad
\epsilon_Q\equiv 1-\frac{Q_{\rm near}}{Q_{\rm ext}}=10^{-4}.
\]
For this value of the charge, the roots of \(f(r)=0\) determine the
inner and outer horizons. We denote \(r_+\) the event horizon and
\(r_-\) the inner horizon. The corresponding surface gravity is
\[
\kappa_+=\frac{1}{2}f'(r_+).
\]

The radial equation can be written 
the radial coordinate ($r$),
\[
f^2(r)\Psi_s''(r)
+
f(r)f'(r)\Psi_s'(r)
+
\left[\omega^2-V_s(r)\right]\Psi_s(r)=0 .
\]
For the purely imaginary near-extremal branch we set
\[
\omega_{\rm NE}=-i\gamma_{\rm NE},
\qquad
\gamma_{\rm NE}>0,
\]
so that the radial equation becomes
\[
f^2(r)\Psi_s''(r)
+
f(r)f'(r)\Psi_s'(r)
-
\left[\gamma^2+V_s(r)\right]\Psi_s(r)=0 .
\]

The QNM boundary conditions are incorporated by factorizing the
radial function as
\[
\Psi_s(r)=e^{h(r;\gamma)}u(r),
\]
where \(u(r)\) is regular in the numerical domain. The factor
\(e^{h}\) contains the ingoing behavior at the event horizon and the
outgoing behavior at infinity. In the implementation we use
\[
H_1(r;\gamma)\equiv h'(r;\gamma)
=
\gamma\left[
R_1(r)-\frac{1}{f'_+(r-r_+)}
+\frac{1}{f'_+r}\right],
\]
and
\begin{eqnarray}
H_2(r;\gamma)&\equiv& h''(r;\gamma)+[h'(r;\gamma)]^2\\
\notag && =
\gamma\left[
R_2(r)+\frac{1}{f'_+(r-r_+)^2}
-\frac{1}{f'_+r^{2}}\right]
+
H_1^2(r;\gamma),
\end{eqnarray}
where
\[
f'_+\equiv f'(r_+),
\]
\[
R_1(r)=1+\frac{2M}{r}
-
\frac{\alpha}{r}\ln\left(\frac{r}{|\alpha|}\right),
\]
and
\[
R_2(r)=R_1'(r)
=
-\frac{2M}{r^2}
-
\frac{\alpha}{r^2}
\left[
1-\ln\left(\frac{r}{|\alpha|}\right)
\right].
\]
Equivalently, up to an irrelevant additive constant,
\[
h(r;\gamma)
=
\gamma\left[
r+2M\ln r
-\frac{\alpha}{2}
\ln^2\left(\frac{r}{|\alpha|}\right)
-\frac{1}{f'_+}\ln\left(\frac{r-r_+}{\bf{r}}\right)
\right].
\]
Near the event horizon this factor behaves as
\[
e^{h}\sim (r-r_+)^{-\gamma/f'_+},
\]
which reproduces the ingoing QNM behavior
\[
\Psi_s\sim e^{-i\omega r_*}=e^{-\gamma r_*},
\qquad r\rightarrow r_+ .
\]
At large radius the same factor reproduces the asymptotic outgoing
behavior of the purely imaginary branch,
\[
\Psi_s\sim e^{+i\omega r_*}=e^{+\gamma r_*},
\qquad r\rightarrow \infty .
\]

Substitution of the factorized ansatz gives the regular differential
equation
\[
f^2(r)u''(r)
+
B(r;\gamma)u'(r)
+
C(r;\gamma)u(r)=0,
\]
with
\[
B(r;\gamma)
=
2f^2(r)H_1(r;\gamma)
+
f(r)f'(r),
\]
and
\[
C(r;\gamma)
=
f^2(r)H_2(r;\gamma)
+
f(r)f'(r)H_1(r;\gamma)
-
\gamma^2
-
V_s(r).
\]

\begin{figure}[H]
\centering
\includegraphics[width=0.4\textwidth]{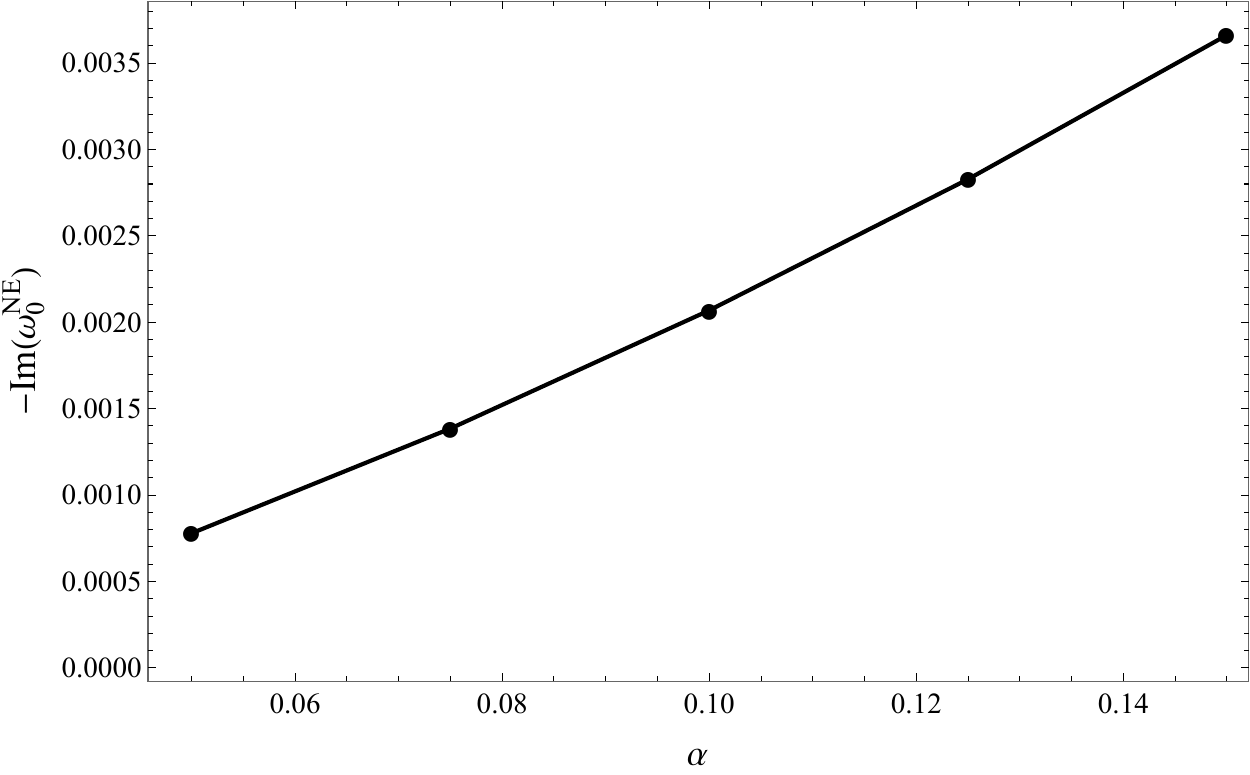}
\caption{
Damping rate $-\mathrm{Im}(\omega_{\rm NE})$ of the
purely imaginary near-extremal mode as a function of the PFDM parameter
\(\alpha\) for the effective gravitational sector $s=2$ with $\ell=2$ and $Q=0.9999Q_{ext}$.  The monotonic increase of $\gamma_{\rm NE}$ indicates that the PFDM parameter enhances the damping of the near-extremal perturbation.
}
\label{fig:gravy_NEAR}
\end{figure}

The equation is solved by a Chebyshev collocation method with a singular-value diagnostic.
The numerical domain is
\[
r\in [r_{\rm in},r_{\rm out}],
\qquad
r_{\rm in}=r_++\epsilon_H,
\qquad
r_{\rm out}=r_{\max}.
\]
We use a Chebyshev-Lobatto grid
\[
x_j=\cos\left(\frac{\pi j}{N}\right),
\qquad j=0,\ldots,N,
\]
mapped to the radial interval as
\[
r_j=
\frac{r_{\rm in}+r_{\rm out}}{2}
+
\frac{r_{\rm out}-r_{\rm in}}{2}x_j .
\]
Let \(D^{(1)}\) be the first Chebyshev differentiation matrix and
\(D^{(2)}=D^{(1)}D^{(1)}\). The discretized equation takes the form
\[
M(\gamma)\mathbf{u}=0,
\]
where
\[
M(\gamma)
=
{\rm diag}(f_j^2)D^{(2)}
+
{\rm diag}(B_j)D^{(1)}
+
{\rm diag}(C_j),
\]
with
\[
f_j=f(r_j),
\qquad
B_j=B(r_j;\gamma),
\qquad
C_j=C(r_j;\gamma).
\]
The rows of \(M(\gamma)\) are normalized in order to improve the
conditioning of the spectral problem. For each trial value of
\(\gamma\), we compute the singular values of \(M(\gamma)\) and define
\[
\sigma_{\rm rel}(\gamma)
=
\frac{\sigma_{\min}[M(\gamma)]}
{\sigma_{\max}[M(\gamma)]}.
\]
A quasinormal frequency is identified by locating an interior minimum
of \(\sigma_{\rm rel}(\gamma)\). Boundary minima of the search interval
are discarded and the search window is shifted and enlarged until an
interior minimum is found. The final near-extremal frequency is then
\[
\omega_{\rm NE}=-i\gamma_{\rm NE}.
\]
Fig.~\ref{fig:gravy_NEAR} shows the damping rate of the purely imaginary near-extremal mode for the effective gravitational sector, $s=2$, versus the PFDM parameter $\alpha$.  In the figure $\gamma_{\rm NE}=-\mathrm{Im}\omega_{\rm NE}$ is observed to increase monotonically with $\alpha$. Thus the PFDM parameter enhances the damping rate of the purely imaginary near-extremal mode and reduces the corresponding relaxation time. Note that the figure shows only the gravitational sector, as the scalar $s=0$ and electromagnetic $s=1$ sectors produce nearly superimposed curves. This implies that the purely imaginary near-extremal branch is almost insensitive to the spin of the perturbing field, and its damping rate is mainly determined by the near-horizon geometry, while spin-dependent contributions to the effective potential are subleading.

A numerical comparison of the purely imaginary near-extremal damping rates for scalar, electromagnetic, and effective axial spin-2 perturbations is shown in Table \ref{tab}. The results show that the three sectors are nearly indistinguishable, with maximum relative differences below $1\%$.
This supports the interpretation that the near-extremal branch is mainly
controlled by the near-horizon geometry rather than by the spin-dependent
structure of the effective potential.
\begin{table}[htbp]
\centering
\begin{tabular}{ccccc}
\hline
\(\alpha\) 
& \(\gamma_{\rm NE}^{(s=0)}\) 
& \(\gamma_{\rm NE}^{(s=1)}\) 
& \(\gamma_{\rm NE}^{(s=2)}\) 
& \(\Delta_{\rm max}(\%)\) \\
\hline
0.050 & 0.00078001 & 0.00078001 & 0.00078481 & 0.612840 \\
0.075 & 0.00138465 & 0.00138464 & 0.00139081 &  0.300860 \\
0.100 & 0.00206866 & 0.00206862 & 0.00207481 & 0.298934 \\
0.125 & 0.00282874 & 0.00282865 & 0.00283681  & 0.288903 \\
0.150 & 0.00366153 & 0.00366138 & 0.00367081 &  0.258147\\
\hline
\end{tabular}
\caption{
Comparison of the near-extremal damping rates 
$\gamma_{\rm NE}=-\mathrm{Im}(\omega_{\rm NE})$ for scalar, electromagnetic,
and effective axial spin-2 perturbations, with $\ell=2$ and $a=0.01$. The quantity 
$\Delta_{\rm max}$ measures the maximum relative difference among the three
sectors.
}
\label{tab}
\end{table}

The appearance of purely imaginary frequencies in the near-extremal
regime can be interpreted as a manifestation of the qualitative change
in the near-horizon geometry as the outer and inner horizons approach
each other. In this limit, the surface gravity decreases and the
Hawking temperature becomes increasingly small, leading to longer
characteristic dynamical timescales. Consequently, the oscillatory
component of the quasinormal spectrum becomes progressively suppressed,
while the perturbations are dominated by slowly damped modes with small
decay rates. This behavior is reminiscent of the near-extremal
Reissner-Nordstr\"om black hole, where a distinct family of purely
imaginary frequencies has been associated with the throat region that
develops near the horizon. The results obtained here suggest that the
Euler-Heisenberg-PFDM geometry exhibits the same qualitative
mechanism: as extremality is approached, the near-horizon sector
progressively decouples from the ordinary photon-sphere branch and
gives rise to a family of modes that encode the dynamics of the throat
geometry rather than the optical properties of the spacetime. We verified that the extracted frequencies remain stable under variations of the spectral resolution N, the horizon offset $\epsilon_H$, and the outer boundary $r_{max}$.

\section{Exact Greybody Factors}
\label{GB}

The effective potentials discussed in the previous sections determine not
only the quasinormal spectrum but also the partial transmission of
incoming waves across the curvature barrier surrounding the black hole.
This phenomenon is encoded in the greybody factor
$\Gamma_s(\omega,\ell)$, which quantifies the probability that a mode of
frequency $\omega$ and angular momentum $\ell$ propagates from the near
horizon region to asymptotic infinity. In the context of Hawking
radiation, greybody factors modify the purely thermal spectrum by
selectively suppressing low-frequency modes and controlling the
efficiency with which different perturbative channels escape the
gravitational potential.

For each spin sector, the radial perturbation equation takes the
Schr\"odinger-like form
\begin{equation}
\frac{d^2\Psi_s}{dr_*^2}
+
\left[
\omega^2
-
V_s(r)
\right]
\Psi_s
=
0,
\label{radial_schrodinger_greybody}
\end{equation}
where the tortoise coordinate is defined by
\begin{equation}
\frac{dr_*}{dr}
=
\frac{1}{f(r)}.
\end{equation}
The boundary conditions appropriate for scattering are
\begin{equation}
\Psi_s \sim
\begin{cases}
T\,e^{-i\omega r_*},
& r_* \to -\infty, \\[1ex]
e^{-i\omega r_*}
+
R\,e^{+i\omega r_*},
& r_* \to +\infty,
\end{cases}
\label{scattering_boundary_conditions}
\end{equation}
where $R$ and $T$ denote the reflection and transmission amplitudes,
respectively. Flux conservation implies
\begin{equation}
|R|^2 + |T|^2 = 1,
\end{equation}
so that the greybody factor is given by
\begin{equation}
\Gamma_s(\omega,\ell)
=
|T(\omega,\ell)|^2.
\label{greybody_definition}
\end{equation}

\begin{figure}[t]
\centering
\includegraphics[width=0.30\textwidth]{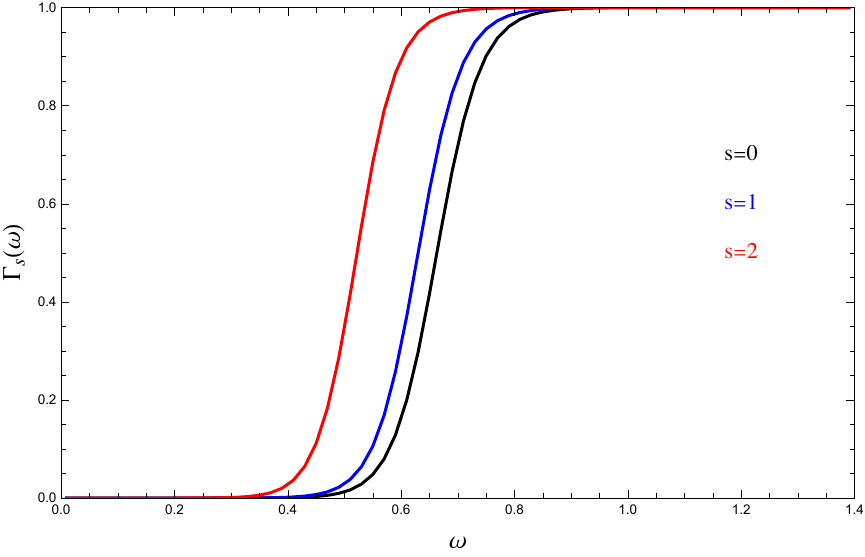}
\caption{
Exact greybody factors for scalar ($s=0$), electromagnetic ($s=1$),
and effective axial spin-2 ($s=2$) perturbations of the
Euler-Heisenberg black hole surrounded by perfect fluid dark matter,
computed by direct numerical integration of the radial wave equation
and asymptotic window fitting. The parameters are
$M=1$, $Q=0.7$, $a=0.01$, $\alpha=0.1$, and $\ell=2$.
The transmission probability increases monotonically with the
frequency and approaches unity in the high-frequency regime. In the
transition region, the effective spin-2 channel is the most transparent,
whereas the scalar channel is the most strongly suppressed.
}
\label{fig:GreybodyFactors}
\end{figure}

To compute $\Gamma_s$, we integrate Eq.~(\ref{radial_schrodinger_greybody})
numerically from a point just outside the event horizon, imposing purely
ingoing boundary conditions. The transmission amplitude is then extracted
by fitting the asymptotic solution over a finite radial window at large
distances. For numerical reproducibility, we now specify the scattering procedure in
more detail. The integration is started at
\begin{equation}
r_{\rm in}=r_h+\epsilon,
\qquad
\epsilon=10^{-7},
\end{equation}
where the near horizon solution is initialized as a purely ingoing mode,
\begin{equation}
\Psi_s(r_{\rm in})
=
e^{-i\omega r_*(r_{\rm in})},
\qquad
\Psi_s'(r_{\rm in})
=
-\frac{i\omega}{f(r_{\rm in})}
e^{-i\omega r_*(r_{\rm in})}.
\end{equation}
The radial equation is then integrated outward up to
\begin{equation}
r_{\rm max}=500M.
\end{equation}
At large radius, the numerical solution is fitted over the asymptotic
window
\begin{equation}
r\in[0.75r_{\rm max},r_{\rm max}]
\end{equation}
to the form
\begin{equation}
\Psi_s(r)
=
A_{\rm in}e^{-i\omega r_*}
+
A_{\rm out}e^{+i\omega r_*}.
\end{equation}
The coefficients \(A_{\rm in}\) and \(A_{\rm out}\) are obtained through
a least squares fit over the full fitting window, rather than by matching
at a single radial point. The greybody factor is then computed as
\begin{equation}
\Gamma_s(\omega,\ell)
=
\frac{1}{|A_{\rm in}|^2}.
\end{equation}
For the representative parameter set used in the scattering and
radiation analysis,
$
M=1,
Q=0.7,
a=0.01,
\alpha=0.1,
$
the event horizon is located at
$
r_h \simeq 1.383046,
$
which yields a Hawking temperature
\begin{equation}
T_H
=
\frac{f'(r_h)}{4\pi}
\simeq 0.046964.
\end{equation}
This value sets the characteristic energy scale of the emission spectra
displayed in Fig.~\ref{fig:AbsorptionEmission}.
This procedure reduces numerical sensitivity to local oscillations in
the asymptotic region and produces smooth transmission curves. The
radial equation was integrated using an adaptive high precision solver
with stiffness switching, working precision 50, and accuracy and
precision goals set to 20. The stability of the results was verified by
varying \(r_{\rm max}\), the fitting window, and the frequency step.
The Pad\'e uncertainty quoted in Table~\ref{tab:QNM_PadeWKB13} is
associated exclusively with the independent quasinormal-mode
calculation and is not propagated into the greybody factors, which are
obtained directly from the numerical solution of the scattering
problem.
Fig.~\ref{fig:GreybodyFactors} displays the exact greybody factors for
scalar ($s=0$), electromagnetic ($s=1$), and effective axial spin-2
($s=2$) perturbations for the representative parameter set
$M=1$, $Q=0.7$, $a=0.01$, $\alpha=0.1$, and $\ell=2$. In all cases, the
transmission probability vanishes in the low-frequency regime, where the
effective potential barrier strongly suppresses outgoing radiation, and
approaches unity at large frequencies, where the barrier becomes
effectively transparent.

A clear hierarchy is observed in the transition region:
\begin{equation}
\Gamma_{2}(\omega)
>
\Gamma_{1}(\omega)
>
\Gamma_{0}(\omega),
\end{equation}
indicating that the effective spin-2 channel becomes transparent at
lower frequencies than the electromagnetic and scalar sectors. This
ordering is a direct consequence of the different heights and widths of
the corresponding effective potentials. Since the scalar potential
reaches the largest maximum, scalar waves experience the strongest
suppression and require higher frequencies to penetrate the barrier.

Transmission starts in the region $\omega \sim 0.4$-$0.7$ which is very close to the characteristic scale $\sqrt{V_0}$ associated with the maxima of the effective potentials.
We then see a direct relation between the scattering problem and the quasinormal analysis of the previous section: the same potential barrier that determines the ringdown spectrum also determines the spectral filtering of the Hawking radiation.

The perfect fluid dark matter parameter $\alpha$ influences the greybody
factors through its impact on the optical structure of the spacetime. As
shown in the preceding sections, increasing $\alpha$ shifts both the
event horizon and the photon sphere inward, raising the characteristic
frequencies $\Omega_c$ and $\lambda_c$. From the scattering perspective,
this translates into a displacement of the transmission curves toward
larger frequencies and a sharper transition between the opaque and
transparent regimes. The Euler-Heisenberg parameter $a$, in contrast,
introduces more localized nonlinear-electrodynamic corrections and
produces comparatively smaller modifications for the parameter range
considered here.

\begin{figure}[h]
\centering
\includegraphics[width=0.3\textwidth]{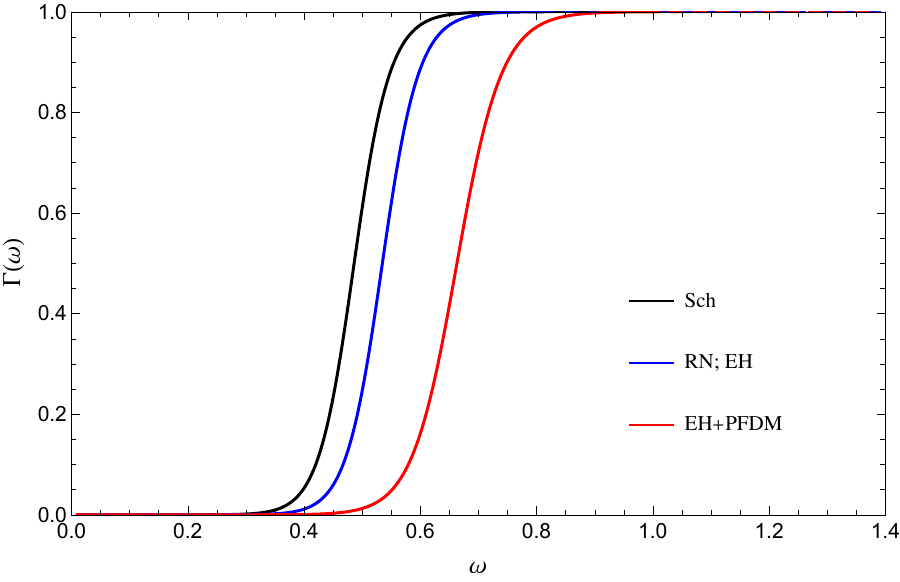}
\includegraphics[width=0.3\textwidth]{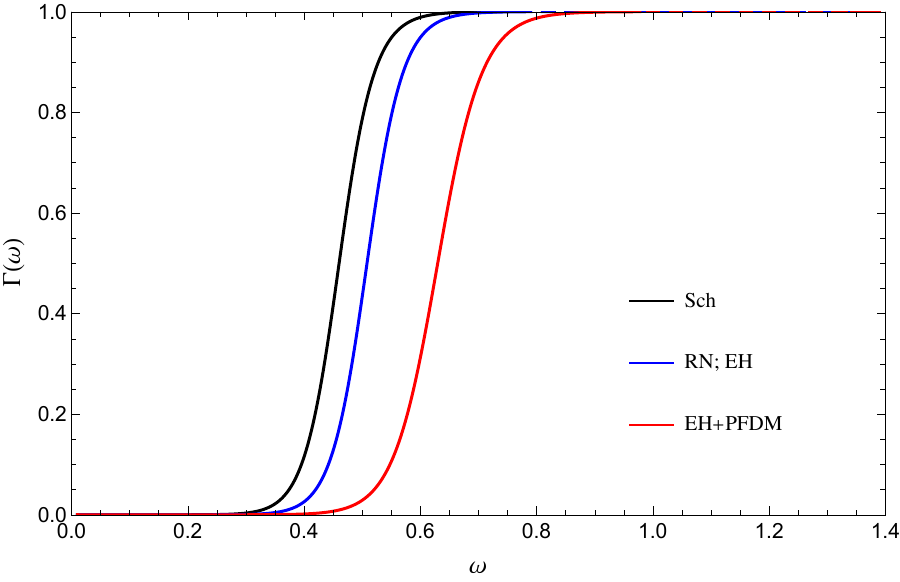}
\includegraphics[width=0.3\textwidth]{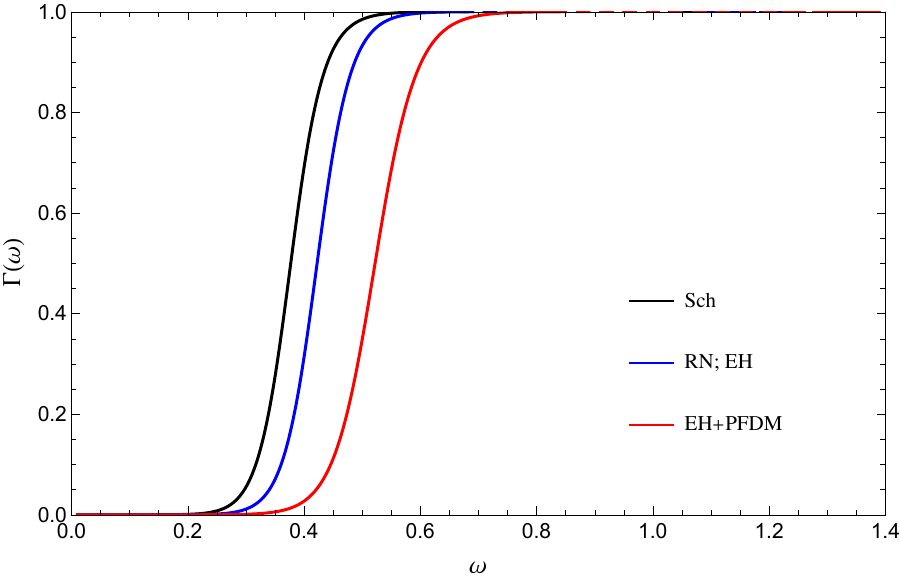}
\caption{Exact greybody factor $\Gamma(\omega)$ as a function of the frequency $\omega$ for scalar (top panel), electromagnetic (central panel), and effective axial spin-2 perturbations (bottom panel) with $\ell=2$. The parameters are fixed as $M=1$, $Q=0.7$, $a=0.01$, and $\alpha=0.1$.}
\label{greyc}
\end{figure}

Fig.~\ref{greyc} displays the greybody factor $\Gamma(\omega)$ for scalar, electromagnetic and effective axial spin-2 perturbations with $\ell=2$, for the geometries of Schwarzschild, RN/EH and EH+PFDM. In all cases the transmission probability shows the usual monotonic behavior, i.e., strongly suppressed in the low-frequency regime and going to unity at sufficiently large frequencies. This is consistent with the observation that the low energy modes are reflected mainly by the effective potential barrier, while the high frequency modes are transmitted almost totally.

The main effect of the charge and the nonlinear electromagnetic corrections is to shift the onset of transmission towards larger frequencies with respect to the Schwarzschild case. However, for the choice of parameters discussed here the RN and Euler-Heisenberg curves are almost indistinguishable. This implies that in this regime the Euler-Heisenberg correction, governed by the parameter $a$, is only a subleading correction to the greybody factor. However, when the PFDM background is included, the curve shifts significantly to the right. Therefore, for a fixed intermediate frequency the EH+PFDM black hole has a smaller transmission probability than the Schwarzschild and RN/EH cases.

This behaviour can be seen as the surrounding PFDM distribution amplifying the effective scattering barrier. The effect is maximum in the transition region, where $\Gamma(\omega)$ varies rapidly from almost zero to almost one, whereas it is negligible in the high-frequency limit, where all curves converge to $\Gamma(\omega)\simeq 1$. In the intermediate frequency range, the same qualitative hierarchy is found for the three perturbation sectors, \begin{equation} \Gamma_{\rm Sch}(\omega)>\Gamma_{\rm RN/EH}(\omega)> \Gamma_{\rm EH+PFDM}(\omega)\,, \end{equation} Thus, the greybody factor is a sensitive probe of the PFDM contribution. The Euler-Heisenberg correction is rather small with respect to the values of the parameters used in Fig.~\ref{greyc}.

\section{Absorption Cross Sections and Hawking Radiation}
\label{ACS}

The greybody factors obtained in the previous section provide the
essential ingredient for determining both the absorption properties and
the Hawking emission spectrum of the Euler-Heisenberg black hole
surrounded by perfect fluid dark matter. While quasinormal modes encode
the characteristic response of the spacetime to perturbations and
greybody factors describe the transmission of waves through the
effective potential barrier, the absorption cross section and the
radiation spectrum quantify the observable consequences of these
processes.

For each perturbative sector, the partial absorption cross section is
defined as
\begin{equation}
\sigma_{\ell}^{(s)}(\omega)
=
\frac{\pi}{\omega^2}
(2\ell+1)\,
\Gamma_s(\omega,\ell).
\label{partial_absorption}
\end{equation}
Summing over all multipoles yields the total absorption cross section,
\begin{equation}
\sigma_{\rm abs}^{(s)}(\omega)
=
\sum_{\ell=s}^{\infty}
\sigma_{\ell}^{(s)}(\omega).
\label{total_absorption}
\end{equation}
In the present analysis we focus on the dominant $\ell=2$ contribution,
which captures the main qualitative features and provides a direct
connection with the quasinormal and scattering results discussed above.

In the high-frequency limit, the absorption cross section approaches the
geometric-optics result determined by the critical impact parameter
$b_c=1/\Omega_c$ associated with the unstable photon sphere,
\begin{equation}
\sigma_{\rm geo}
=
\pi b_c^2
=
\frac{\pi}{\Omega_c^2}.
\label{geometric_cross_section}
\end{equation}
This relation highlights once again the central role of the optical
structure: the same null geodesics that govern the eikonal quasinormal
spectrum also determine the asymptotic absorption properties of the
black hole.

The Hawking emission spectrum is given by
\begin{equation}
\frac{d^2E}{d\omega\,dt}
=
\frac{1}{2\pi}
\sum_{\ell}
\frac{
(2\ell+1)\,
\omega\,
\Gamma_s(\omega,\ell)
}{
\exp(\omega/T_H)-1
},
\label{hawking_spectrum}
\end{equation}
where $T_H$ is the Hawking temperature,
\begin{equation}
T_H
=
\frac{f'(r_h)}{4\pi}.
\label{hawking_temperature}
\end{equation}
The greybody factor acts as a spectral filter, suppressing low-frequency
modes and reshaping the ideal blackbody distribution. Consequently, the
emission spectrum reflects a nontrivial interplay between the horizon
temperature and the transmission properties of the effective potential.

\begin{figure}[h]
\centering
\includegraphics[width=0.3\textwidth]{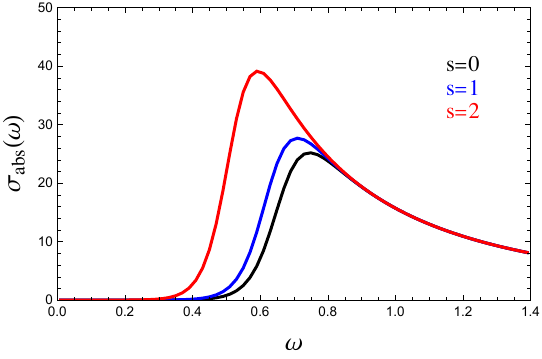}\\
\includegraphics[width=0.33\textwidth]{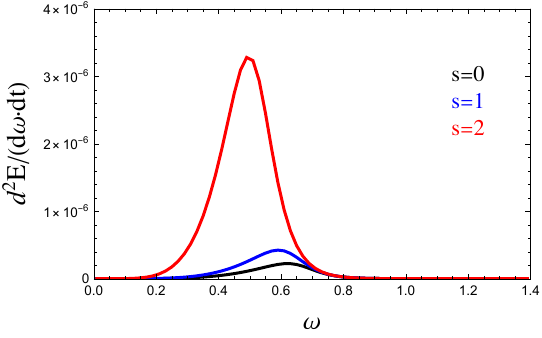}
\caption{
Absorption cross sections (top panel) and Hawking energy emission rates (bottom panel) for
scalar ($s=0$), electromagnetic ($s=1$), and effective axial spin-2
($s=2$) perturbations of the Euler-Heisenberg black hole surrounded by
perfect fluid dark matter. The calculations are based on the exact
greybody factors obtained by direct numerical integration of the radial
wave equation for the representative parameter set
$M=1$, $Q=0.7$, $a=0.01$, $\alpha=0.1$, and $\ell=2$.
Top panel displays the partial absorption cross sections, which rise
rapidly once the transmission probability becomes appreciable and
approach the geometric-optics regime at high frequencies.
Bottom panel shows the corresponding Hawking energy emission spectra,
which exhibit well-defined peaks resulting from the competition between
the Bose-Einstein occupation factor and greybody suppression.
The effective spin-2 channel displays the largest low- and
intermediate-frequency absorption, while the PFDM halo shifts the
characteristic emission toward higher frequencies and modifies the
overall radiative efficiency.
}
\label{fig:AbsorptionEmission}
\end{figure}

Fig.~\ref{fig:AbsorptionEmission} presents the absorption cross
sections and the corresponding Hawking energy emission rates for the
same representative configuration used throughout this work,
$M=1$, $Q=0.7$, $a=0.01$, $\alpha=0.1$, and $\ell=2$. The absorption
cross sections increase rapidly once the greybody factors become
appreciable and gradually approach the geometric-optics limit. The
ordering observed in the greybody factors is preserved, with the
effective spin-2 channel displaying the largest low- and
intermediate-frequency absorption.

The Hawking spectra exhibit a pronounced peak at frequencies determined
by the competition between the Bose-Einstein factor and the greybody
suppression. At very low frequencies, the transmission probability is
strongly suppressed, while at high frequencies the thermal occupation
factor decreases exponentially. The resulting spectra therefore possess
well-defined maxima that provide the dominant contribution to the total
radiated power.

The perfect fluid dark matter parameter $\alpha$ affects the emission
process in two complementary ways. First, by shifting the horizon inward
it modifies the Hawking temperature through Eq.~(\ref{hawking_temperature}).
Second, by altering the effective potential it changes the transmission
probabilities and the absorption cross sections. In the parameter range
considered here, these effects combine to shift the emission peak toward
higher frequencies and to enhance the overall luminosity. The
Euler-Heisenberg parameter $a$ introduces smaller but systematic
corrections associated with nonlinear electrodynamic effects near the
event horizon.

The absorption and emission analyses demonstrate that
the dark-matter environment leaves measurable signatures not only in the
ringdown spectrum but also in the thermodynamic and scattering
properties of the black hole. The PFDM halo modifies the effective
spectral window through which Hawking radiation escapes, thereby
providing an additional observational channel through which the
influence of dark matter and nonlinear electrodynamics may be probed.

\section{Neutrino-antineutrino annihilation in the Euler-Heisenberg-PFDM geometry}
\label{sec:nu_nubar_eh_pfdm}

In this section, we evaluate the relativistic enhancement of the energy deposition rate generated by
the annihilation channel \cite{Salmonson:1999es,Asano:2000ib,Asano:2000dq,Lambiase:2020iul,Lambiase:2020pkc,Pantig:2025eda,Shi:2023kid,Mannobova:2026jna,Kholmuminov:2026vpv,Alloqulov:2024sns}
\begin{equation}
\nu+\bar{\nu}\longrightarrow e^-+e^+ ,
\end{equation}
outside a static Euler-Heisenberg black hole surrounded by perfect-fluid dark matter (PFDM).
The calculation follows the standard neutrino-sphere treatment, but with the redshift and bending
factors computed from the Euler-Heisenberg-PFDM lapse function.

For later numerical work, it is convenient to introduce dimensionless variables
\begin{equation}
\rho=\frac{r}{M},\qquad
\rho_R=\frac{R}{M},\qquad
q=\frac{Q}{M},\qquad
\eta=\frac{a}{M^2},\qquad
\lambda=\frac{\alpha}{M}.
\label{eq:dimensionless_variables_ehpfdm}
\end{equation}
In these variables, the lapse \eqref{eq:fmetric} becomes
\begin{equation}
{\cal F}(\rho)
=
1-\frac{2}{\rho}
+\frac{q^2}{\rho^2}
-\frac{\eta q^4}{20\rho^6}
+\frac{\lambda}{\rho}\ln\!\left(\frac{\rho}{|\lambda|}\right),
\label{eq:dimensionless_lapse_ehpfdm}
\end{equation}
with the smooth limiting prescription \(\lambda\ln(\rho/|\lambda|)\rightarrow 0\) as
\(\lambda\rightarrow0\).

The local annihilation energy deposition rate per unit volume may be written as
\begin{equation}
\frac{dE}{dt\,dV}
=
2K G_F^2 F(r)
\int_0^\infty\!\!\int_0^\infty
n(\epsilon_\nu)n(\epsilon_{\bar\nu})
(\epsilon_\nu+\epsilon_{\bar\nu})
\epsilon_\nu^3\epsilon_{\bar\nu}^3
\,d\epsilon_\nu\,d\epsilon_{\bar\nu}.
\label{eq:local_annihilation_integral}
\end{equation}
The coefficient \(K\) contains the weak-interaction angular factors.  For electron neutrinos one has
\begin{equation}
K_{\nu_e\bar\nu_e}
=
\frac{1}{6\pi}
\left(
1+4\sin^2\theta_W+8\sin^4\theta_W
\right),
\label{eq:K_electron_neutrino}
\end{equation}
while for \(\nu_\mu\bar\nu_\mu\) or \(\nu_\tau\bar\nu_\tau\)
\begin{equation}
K_{\nu_\mu\bar\nu_\mu}
=
K_{\nu_\tau\bar\nu_\tau}
=
\frac{1}{6\pi}
\left(
1-4\sin^2\theta_W+8\sin^4\theta_W
\right).
\label{eq:K_mu_tau_neutrino}
\end{equation}
The purely geometrical part is encoded in
\begin{equation}
F(r)=
\int\!\!\int
\left(1-\boldsymbol{\Omega}_{\nu}\cdot\boldsymbol{\Omega}_{\bar{\nu}}\right)^2
d\Omega_{\nu}d\Omega_{\bar{\nu}}
=
\frac{2\pi^2}{3}
(1-x)^4(x^2+4x+5),
\label{eq:angular_factor}
\end{equation}
where \(x\) determines the local opening angle of the neutrino trajectories.

Assuming a Fermi-Dirac distribution at local temperature \(T(r)\),
\begin{equation}
n(\epsilon_\nu)=
\frac{2}{h^3}
\frac{1}{\exp[\epsilon_\nu/(kT)]+1},
\label{eq:fermi_dirac}
\end{equation}
the energy integral gives the characteristic \(T^9\) dependence
\begin{equation}
\frac{dE}{dt\,dV}
=
\frac{21\zeta(5)\pi^4}{h^6}
K G_F^2 F(r)\,[kT(r)]^9 .
\label{eq:local_rate_T9}
\end{equation}
This strong ninth-power dependence makes the result highly sensitive to gravitational redshift.

For a static observer in the metric \eqref{eq:dimensionless_lapse_ehpfdm}, Tolman's relation gives
\begin{equation}
T(r)\sqrt{f(r)}=T(R)\sqrt{f(R)} ,
\label{eq:tolman_relation_ehpfdm}
\end{equation}
where \(R\) is the neutrino-sphere radius.  Hence the local temperature increases relative to the
temperature at infinity whenever the lapse \(f(r)\) is small.

The null-geodesic impact parameter \(b=L/E\) satisfies
\begin{equation}
b^2=\frac{r^2\sin^2\vartheta}{f(r)} .
\label{eq:impact_parameter_general}
\end{equation}
For the limiting ray emitted tangentially from the neutrino sphere, \(b^2=R^2/f(R)\).  Therefore
the angular variable appearing in \eqref{eq:angular_factor} is
\begin{equation}
x^2
=
1-\frac{R^2}{r^2}\frac{f(r)}{f(R)} .
\label{eq:x_general_ehpfdm}
\end{equation}
In dimensionless form, \(r=Ry\), this becomes
\begin{equation}
X^2(y;\rho_R,q,\eta,\lambda)
=
1-
\frac{1}{y^2}
\frac{{\cal F}(\rho_R y)}
{{\cal F}(\rho_R)} .
\label{eq:X_dimensionless}
\end{equation}
This equation contains the full geometrical effect of charge, nonlinear electrodynamics, and the
PFDM halo on the local collision angle of neutrinos and antineutrinos.

The total deposited power is obtained by integrating the local rate over the proper spatial volume.
For the metric \eqref{eq:dimensionless_lapse_ehpfdm}, the proper radial volume factor is
\(\sqrt{g_{rr}}=1/\sqrt{f(r)}\).  After using the redshift relation, the Newtonian-normalized
deposition rate is
\begin{equation}
\frac{\dot Q}{\dot Q_{\rm Newt}}
=
3\,[f(R)]^{9/4}
\int_1^\infty
(1-X)^4(X^2+4X+5)\,
\frac{y^2\,dy}{[f(Ry)]^5}.
\label{eq:qdot_ratio_general_ehpfdm}
\end{equation}
Finally, substituting \eqref{eq:dimensionless_lapse_ehpfdm} gives the working expression
\begin{equation}
\frac{\dot Q}{\dot Q_{\rm Newt}}
=
3\,[{\cal F}(\rho_R)]^{9/4}
\int_1^\infty
\frac{
(1-X)^4(X^2+4X+5)y^2
}{
[{\cal F}(\rho_R y)]^5
}\,dy
\label{eq:qdot_ratio_dimensionless_ehpfdm}
\end{equation}
with \(X\) given by Eq.~\eqref{eq:X_dimensionless}.  This is the master equation used in the
figures below.

\begin{figure}[t]
\centering
\includegraphics[width=1\linewidth]{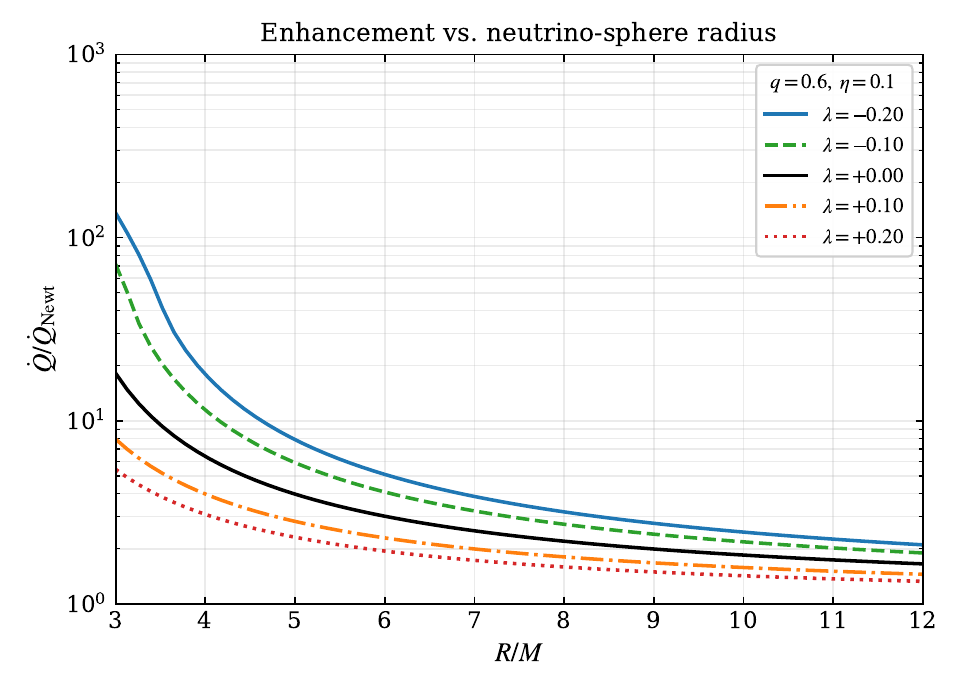}
\caption{
Newtonian-normalized neutrino-annihilation energy deposition rate
\(\dot Q/\dot Q_{\rm Newt}\) as a function of the neutrino-sphere radius \(R/M\) for the
Euler-Heisenberg-PFDM geometry.  We use \(q=0.6\), \(\eta=0.1\), and vary the PFDM
parameter \(\lambda=\alpha/M\).  Negative \(\lambda\) lowers the lapse near the emitting
surface and enhances both the gravitational redshift and the neutrino bending contribution,
thereby increasing the deposited power.  Positive \(\lambda\) acts in the opposite direction for
the displayed range.}
\label{fig:ehpfdm_qdot_vs_R}
\end{figure}

\begin{figure}[t]
\centering
\includegraphics[width=1\linewidth]{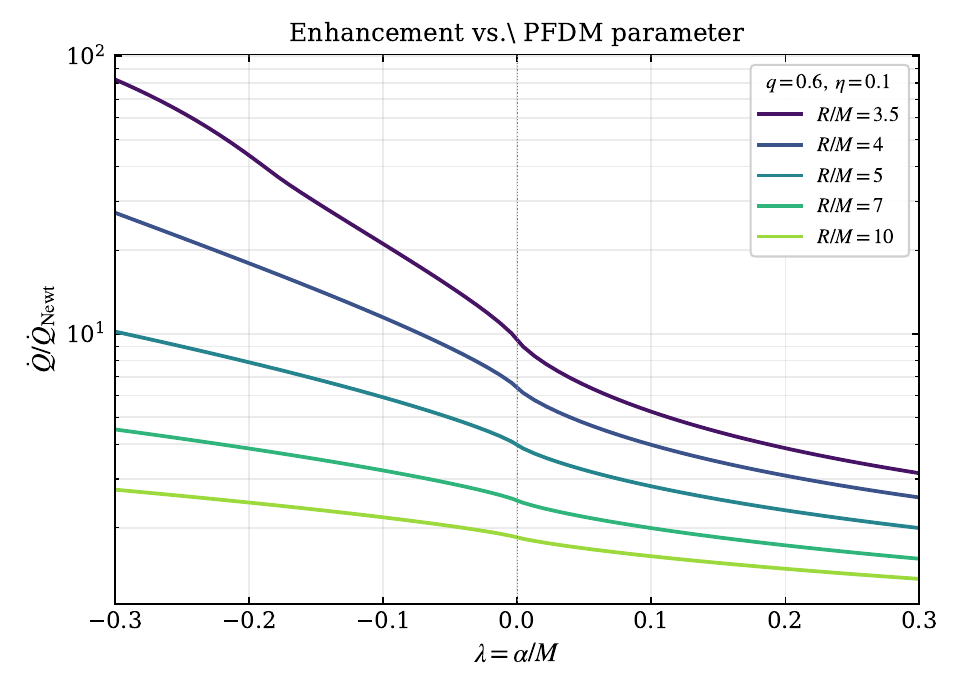}
\caption{
Dependence of \(\dot Q/\dot Q_{\rm Newt}\) on the PFDM parameter
\(\lambda=\alpha/M\) for several values of \(R/M\), with \(q=0.6\) and \(\eta=0.1\).
The enhancement is strongest for the smallest neutrino-sphere radius because the annihilation
rate carries high powers of the redshifted temperature and the lapse function.  The monotonic
decrease with increasing \(\lambda\) in this calculations follows from the corresponding increase of
\({\cal F}(R)\), which reduces the relativistic focusing contribution.}
\label{fig:ehpfdm_qdot_vs_alpha}
\end{figure}

For the numerical calculations used in Figs.~\ref{fig:ehpfdm_qdot_vs_R} and
\ref{fig:ehpfdm_qdot_vs_alpha}, the Euler-Heisenberg term is numerically subdominant at
\(R/M\gtrsim4\) because it scales as \(r^{-6}\).  Its contribution can become appreciable only for
smaller neutrino-sphere radii, larger charge, or larger \(\eta=a/M^2\).  By contrast, the PFDM term
has a longer-range radial dependence and can noticeably modify the deposited power even for
moderately large \(R/M\).
The master expression~\eqref{eq:qdot_ratio_dimensionless_ehpfdm} is evaluated
by adaptive quadrature in the dimensionless variable \(y=r/R\). The integrand
is regular at the lower limit since \(X(1)=0\) and the prefactor
\((1-X)^{4}(X^{2}+4X+5)\rightarrow 5\) there, while the integrand decays as
\((1-X)^{4}\sim y^{-4}\) at \(y\rightarrow\infty\) for any fixed \(\rho_{R}\)
outside the outer horizon. We use a relative tolerance \(10^{-7}\) and split
the integral at \(y_{\max}=400\) with a tail correction; convergence has been
verified by doubling both bounds. The implementation has been cross-checked
against three independent limits:

(i) the Schwarzschild geometry (\(q=\eta=\lambda=0\)) reproduces the
Salmonson-Wilson framework, \(\dot Q/\dot Q_{\mathrm{Newt}}\!\approx\!
28.8,\,7.33,\,3.15,\,1.87\) for \(R/M=3,4,6,10\);

(ii) the Reissner-Nordstr\"om limit (\(\eta=\lambda=0\)) recovers the values
previously reported for charged black holes;

(iii) the EH-RN sub-limit (\(\lambda=0\)) agrees with the RN value to within
\(10^{-3}\) for our fiducial \((q,\eta)=(0.6,0.1)\), in line with the analytic
estimate \(\eta q^{4}/(20\rho^{6})\!\sim\!10^{-6}\) at \(\rho=3\).

\begin{figure}[t]
 \centering
 \includegraphics[width=1\linewidth]{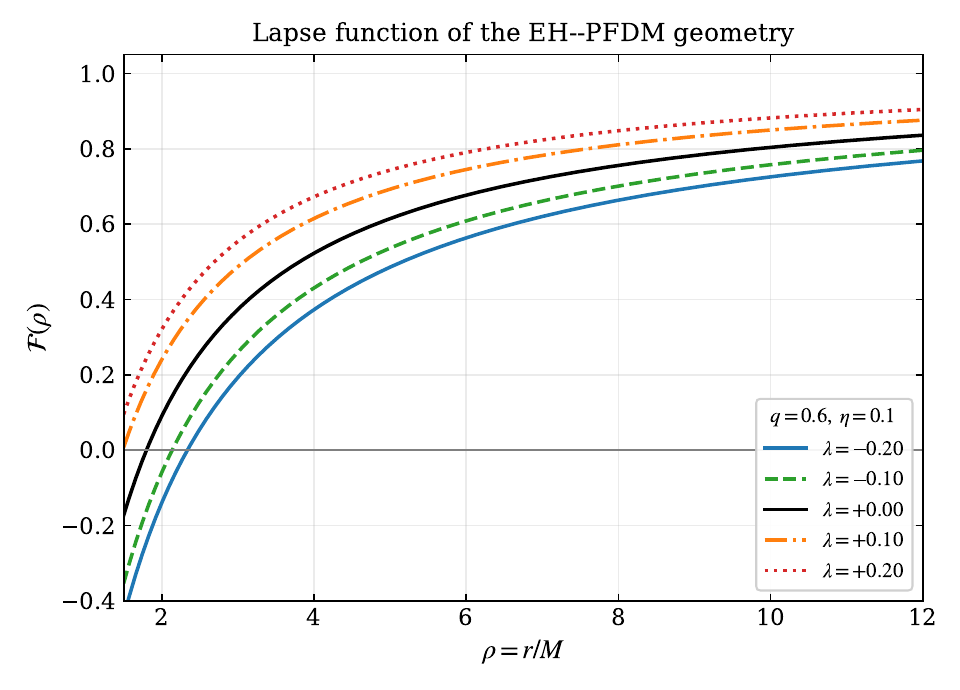}
 \caption{Lapse function $\mathcal{F}(\rho)$ of the EH-PFDM geometry for $q=0.6$,
 $\eta=0.1$, and several values of $\lambda$. Negative $\lambda$ shifts the outer horizon
 outward and depresses $\mathcal{F}$ across the exterior region, providing the geometric
 origin of the enhancement seen in
 Figs.~\ref{fig:ehpfdm_qdot_vs_R}-\ref{fig:ehpfdm_qdot_vs_alpha}.}
 \label{fig:ehpfdm_lapse}
 \end{figure}

These cross-checks are summarised in Table~\ref{tab:limits}. Fig.~\ref{fig:ehpfdm_qdot_vs_R} shows that the Newtonian-normalised
deposition rate falls off rapidly with \(R/M\); this reflects the steep
\(T^{9}\) Fermi-Dirac dependence of the local annihilation
kernel~\eqref{eq:local_rate_T9} combined with the \([\mathcal{F}(R)]^{9/4}\)
Tolman factor. At \(R/M=3\) the deviation from flat space is already by more
than an order of magnitude even in the pure Schwarzschild case. Switching on
a negative PFDM coefficient further lowers \(\mathcal{F}(R)\) (see the
auxiliary Fig.~\ref{fig:ehpfdm_lapse}), amplifying both the redshifted
temperature and the gravitational focusing of the neutrino trajectories. The
cumulative effect is striking: at \(\lambda=-0.2\),
\(\dot Q/\dot Q_{\mathrm{Newt}}\) at \(R/M=3\) reaches values close to
\(1.4\times10^{2}\). Positive \(\lambda\), by contrast, raises
\(\mathcal{F}\) and monotonically reduces the deposited power, with the
curves remaining strictly above unity at all displayed radii.

Fig.~\ref{fig:ehpfdm_qdot_vs_alpha} repeats the analysis at fixed \(R/M\)
and varying \(\lambda\). All curves are monotonically decreasing in
\(\lambda\), consistent with the trend in
Fig.~\ref{fig:ehpfdm_qdot_vs_R}. The slope is steepest at the smallest
\(R/M\): for \(R/M=3.5\) a swing of \(\Delta\lambda=0.2\) changes the
enhancement by roughly a decade, whereas at \(R/M=10\) the corresponding
change is at most a factor of \(\sim\!1.7\). This is the announced
consequence of the \(T^{9}\) sensitivity: small, long-range modifications of
the lapse close to the emitting surface are exponentiated into very large
modifications of the deposited power.

\begin{figure}[t]
 \centering
 \includegraphics[width=1\linewidth]{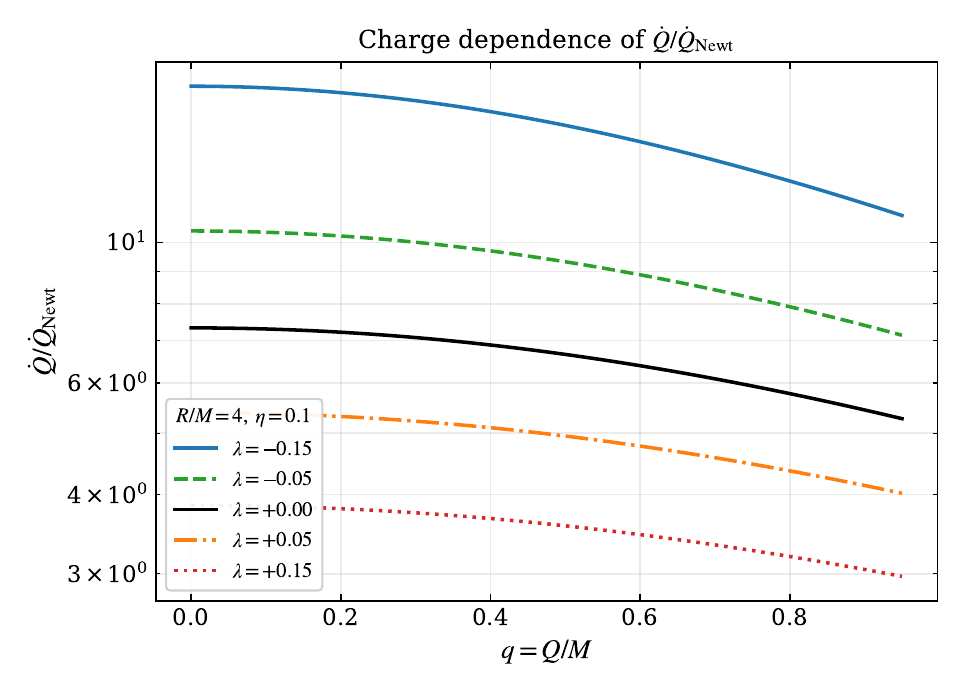}
 \caption{Charge dependence of the Newtonian-normalised deposition rate
 $\dot Q/\dot Q_{\mathrm{Newt}}$ at fixed $R/M=4$, $\eta=0.1$. The Coulomb term raises
 the lapse and produces a mild monotone decrease with $q$. The role of the PFDM
 parameter $\lambda$ dominates over the entire range.}
 \label{fig:ehpfdm_qdot_vs_q}
 \end{figure}

Fig.~\ref{fig:ehpfdm_qdot_vs_q} isolates the dependence on the charge
parameter \(q\) at fixed \(R/M=4\), \(\eta=0.1\), and five values of
\(\lambda\). Increasing \(q\) raises \(\mathcal{F}\) through the
\(+q^{2}/\rho^{2}\) Coulomb contribution and therefore \emph{reduces}
\(\dot Q/\dot Q_{\mathrm{Newt}}\). The reduction is monotonic and weak
(\(\lesssim 40\%\)) over \(0\le q\le 0.95\), to be contrasted with the
dramatic sensitivity to \(\lambda\) at the same radius. Charge therefore
acts as a screening effect on the PFDM-driven enhancement rather than as
an independent amplifier.

\begin{figure}[t]
\centering
\includegraphics[width=\columnwidth]{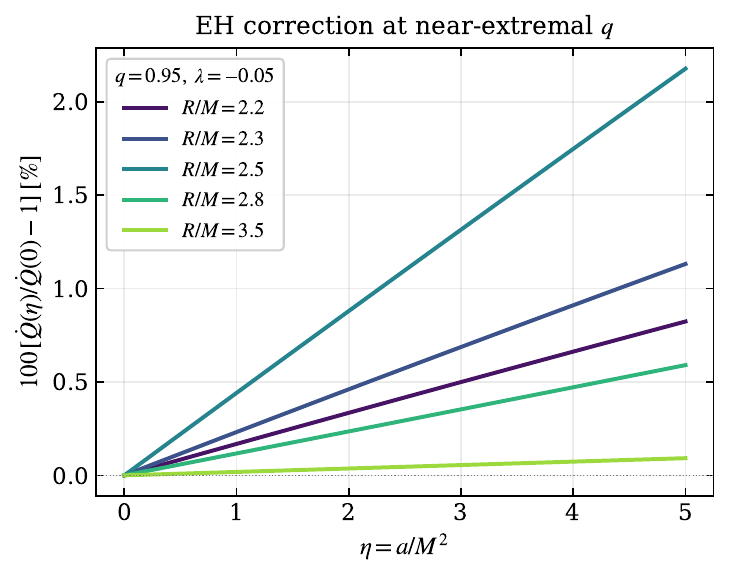}
\caption{Relative deviation
$100\,[\dot Q(\eta)/\dot Q(0)-1]$ induced by the
Euler-Heisenberg correction at near-extremal charge $q=0.95$ and
$\lambda=-0.05$, for several values of the neutrino-sphere radius
$R/M$. The QED contribution is positive and linear in $\eta$, and is
largest when the emission surface lies just outside the outer horizon.
At $R/M=3.5$ the EH effect is below $0.2\%$ across the displayed
range.}
\label{fig:ehpfdm_qdot_vs_eta_R}
\end{figure}

\begin{figure}[t]
\centering
\includegraphics[width=\columnwidth]{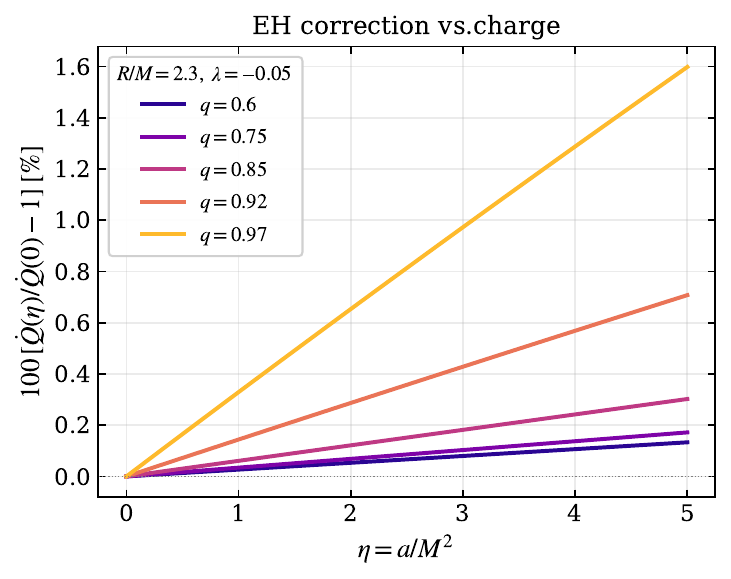}
\caption{Relative deviation
$100\,[\dot Q(\eta)/\dot Q(0)-1]$ induced by the Euler-Heisenberg
correction at fixed near-horizon radius $R/M=2.3$ and $\lambda=-0.05$,
for several values of the charge $q$. The effect grows monotonically
with $q$ but remains below $\sim\!1.6\%$ for $\eta\leq 5$, confirming
that the QED contribution is a small correction within the parameter
range usually considered.}
\label{fig:ehpfdm_qdot_vs_eta_q}
\end{figure}

The Euler-Heisenberg sector is examined in
Figs.~\ref{fig:ehpfdm_qdot_vs_eta_R} and~\ref{fig:ehpfdm_qdot_vs_eta_q}.
Because the EH term in the lapse, \(-\eta q^{4}/(20\rho^{6})\), is suppressed
by six powers of the radial coordinate, the direct dependence of
\(\dot Q/\dot Q_{\mathrm{Newt}}\) on \(\eta\) is too small to be visible on
a linear scale. We therefore plot the relative deviation
\(100\,[\dot Q(\eta)/\dot Q(0)-1]\) in two complementary configurations.
Fig.~\ref{fig:ehpfdm_qdot_vs_eta_R} fixes near-extremal charge \(q=0.95\)
and varies the neutrino-sphere radius near the horizon: the EH contribution
is largest just outside the horizon (\(R/M=2.2\)-\(2.5\)), reaching about
\(2\%\) at \(\eta=5\), and decreases rapidly with \(R/M\), falling below
\(0.2\%\) already at \(R/M=3.5\). Fig.~\ref{fig:ehpfdm_qdot_vs_eta_q}
fixes \(R/M=2.3\) and varies the charge: the deviation grows monotonically
with \(q\), passing from \(\sim\!0.1\%\) at \(q=0.6\) to \(\sim\!1.6\%\) at
\(q=0.97\), again at \(\eta=5\). In both panels the QED correction is a
linear, positive function of \(\eta\), never exceeding a few percent within
the parameter range typically considered in the literature on EH black
holes. The positivity follows from the sign of the EH term, which lowers
\(\mathcal{F}\) for \(\eta>0\) and thereby strengthens redshift and lensing.
We conclude that, at the level of the
\(\nu\bar\nu\rightarrow e^{-}e^{+}\) channel, the QED correction is
phenomenologically negligible compared with the PFDM contribution and
competes with it only in a narrow strip of parameter space close to the
horizon and at near-extremal charge.

\begin{figure}[t]
 \centering
 \includegraphics[width=1\linewidth]{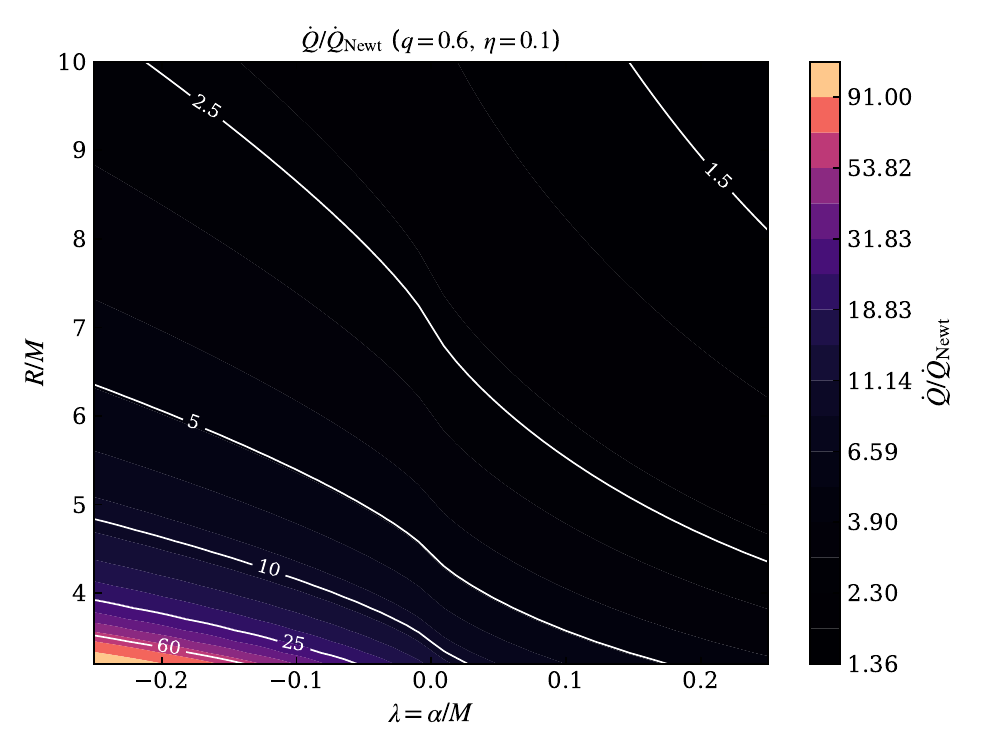}
 \caption{Contour plot of $\dot Q/\dot Q_{\mathrm{Newt}}$ in the $(\lambda,R/M)$ plane
 for $(q,\eta)=(0.6,0.1)$. The white contours mark constant enhancement factors. The
 order-of-magnitude band lies in the negative-$\lambda$, small-$R$ corner relevant for
 GRB central-engine applications.}
 \label{fig:ehpfdm_qdot_contour}
 \end{figure}

Fig.~\ref{fig:ehpfdm_qdot_contour} consolidates these trends in the
\((\lambda,R/M)\) plane at fixed \((q,\eta)=(0.6,0.1)\). The contours of
constant \(\dot Q/\dot Q_{\mathrm{Newt}}\) are nearly straight in
\(\log\)-scale and tilt strongly toward the negative-\(\lambda\) corner.
The region of order-of-magnitude relativistic enhancement
(\(\dot Q/\dot Q_{\mathrm{Newt}}>10\)) is bounded by a curve that passes
approximately through \((\lambda,R/M)=(-0.05,3.5)\) and \((-0.2,5)\), which
is the physically interesting band for gamma-ray burst central-engine
applications.
Table~\ref{tab:qdot_R_lam} provides
\(\dot Q/\dot Q_{\mathrm{Newt}}\) on the fiducial grid, allowing readers to
interpolate directly to other neutrino-sphere radii without rerunning the
code. Table~\ref{tab:limits} contains the limiting-case calculations
discussed above; the identity of the Schwarzschild row with the standard
result, and of the EH-RN row with the corresponding RN entry at the
displayed precision, serves as the principal numerical verification of the
calculation. Table~\ref{tab:relative} quantifies the relative shift induced
by the PFDM parameter: for \(\lambda=-0.10\) the enhancement at \(R/M=4\)
is already \(\sim\!80\%\) above the Reissner-Nordstr\"om-Euler-Heisenberg
baseline, and for \(\lambda=-0.20\) the shift exceeds \(180\%\). Equivalent
positive-\(\lambda\) cases reduce the rate by up to \(\sim\!52\%\). These
numbers establish that a perfect-fluid dark-matter halo surrounding a
charged Euler-Heisenberg black hole is a quantitatively significant
ingredient of the gravitational amplification of neutrino-pair annihilation,
and that the dominant astrophysical degree of freedom controlling the
deposited energy is the PFDM coefficient \(\lambda=\alpha/M\) rather than
the QED parameter \(\eta=a/M^{2}\).

\begingroup
\setlength{\tabcolsep}{4pt}

\begin{table*}[t]
\caption{\label{tab:qdot_R_lam}%
Newtonian-normalized neutrino-annihilation deposition rate
$\dot Q/\dot Q_{\rm Newt}$ for the Euler-Heisenberg-PFDM geometry,
with $q=0.6$ and $\eta=0.1$. Rows label the neutrino-sphere radius
$R/M$; columns label the PFDM parameter $\lambda=\alpha/M$.}
\begin{ruledtabular}
\begin{tabular}{cccccccc}
$R/M$ & $\lambda=-0.20$ & $\lambda=-0.10$ & $\lambda=-0.05$ & $\lambda=0$ & $\lambda=+0.05$ & $\lambda=+0.10$ & $\lambda=+0.20$\\ \hline
$3.0$  & $136.68$ & $71.69$ & $39.43$ & $18.20$ & $10.66$ & $7.94$ & $5.43$\\
$3.5$  & $43.90$  & $21.07$ & $14.78$ & $9.50$  & $6.54$  & $5.24$ & $3.87$\\
$4.0$  & $17.94$  & $11.45$ & $8.89$  & $6.39$  & $4.77$  & $3.98$ & $3.08$\\
$5.0$  & $7.89$   & $5.90$  & $4.98$  & $3.98$  & $3.23$  & $2.82$ & $2.31$\\
$6.0$  & $5.10$   & $4.08$  & $3.58$  & $3.01$  & $2.56$  & $2.29$ & $1.94$\\
$8.0$  & $3.18$   & $2.72$  & $2.48$  & $2.20$  & $1.96$  & $1.81$ & $1.59$\\
$10.0$ & $2.47$   & $2.18$  & $2.03$  & $1.85$  & $1.69$  & $1.58$ & $1.42$\\
\end{tabular}
\end{ruledtabular}
\end{table*}

\begin{table*}[t]
\caption{\label{tab:limits}%
Reduction to limiting geometries. Each entry is
$\dot Q/\dot Q_{\rm Newt}$ at the indicated $R/M$. The Schwarzschild
row reproduces the Salmonson-Wilson calculation; the EH-RN row agrees
with the RN entry at this precision because the Euler-Heisenberg
contribution is $r^{-6}$-suppressed at moderate $R/M$ and $q$.}
\begin{ruledtabular}
\begin{tabular}{lcccc}
Geometry & $R/M=3$ & $R/M=4$ & $R/M=6$ & $R/M=10$\\ \hline
Schwarzschild                              & $28.82$  & $7.33$  & $3.15$ & $1.87$\\
RN $(q=0.6)$                               & $18.20$  & $6.39$  & $3.01$ & $1.85$\\
EH-RN $(q=0.6,\;\eta=0.1)$                & $18.20$  & $6.39$  & $3.01$ & $1.85$\\
PFDM only $(\lambda=-0.10)$                & $101.67$ & $13.70$ & $4.29$ & $2.21$\\
PFDM only $(\lambda=+0.10)$                & $10.50$  & $4.46$  & $2.39$ & $1.60$\\
EH-PFDM $(q=0.6,\;\eta=0.1,\;\lambda=-0.10)$ & $71.69$  & $11.45$ & $4.08$ & $2.18$\\
EH-PFDM $(q=0.6,\;\eta=0.1,\;\lambda=+0.10)$ & $7.94$   & $3.98$  & $2.29$ & $1.58$\\
\end{tabular}
\end{ruledtabular}
\end{table*}

\begin{table*}[t]
\caption{\label{tab:relative}%
Percentage change in $\dot Q/\dot Q_{\rm Newt}$ relative to the
$\lambda=0$ baseline ($q=0.6$, $\eta=0.1$). Negative $\lambda$
produces strong amplification, particularly at small $R/M$ where the
redshifted Fermi-Dirac temperature dominates.}
\begin{ruledtabular}
\begin{tabular}{ccccc}
$\lambda$ & $R/M=3$ & $R/M=4$ & $R/M=6$ & $R/M=10$\\ \hline
$-0.20$ & $+651.2\%$ & $+180.7\%$ & $+69.2\%$ & $+33.5\%$\\
$-0.10$ & $+294.0\%$ & $+79.2\%$  & $+35.6\%$ & $+18.1\%$\\
$-0.05$ & $+116.7\%$ & $+39.1\%$  & $+19.0\%$ & $+9.9\%$\\
$+0.05$ & $-41.4\%$  & $-25.3\%$  & $-15.0\%$ & $-8.7\%$\\
$+0.10$ & $-56.4\%$  & $-37.7\%$  & $-23.9\%$ & $-14.5\%$\\
$+0.20$ & $-70.2\%$  & $-51.7\%$  & $-35.5\%$ & $-22.9\%$\\
\end{tabular}
\end{ruledtabular}
\end{table*}

\endgroup

\section{Conclusions}
\label{conclusions}
We have studied the dynamical, scattering and radiative properties of Euler–Heisenberg black holes surrounded by perfect fluid dark matter (EH+PFDM). Our aim was to establish a unified framework connecting photon-sphere dynamics, quasinormal ringing, greybody factors, Hawking radiation and neutrino energy deposition in the same spacetime geometry.

The EH+PFDM black hole is a superposition of two physically different modifications of the Reissner-Nordstr\"om geometry: a compact near-horizon correction due to nonlinear electrodynamics and an extended logarithmic one related to the dark matter environment. In our analysis we find that these two sectors leave qualitatively different signatures in observables. The dominant effect is due to the PFDM parameter $\alpha$ which moves the event horizon and photon sphere inward, while the Euler-Heisenberg coupling $a$ gives rise to subleading corrections which are important only for sufficiently large electric charge.

We computed quasinormal frequencies for scalar, electromagnetic and effective axial spin-2 perturbations using the thirteenth-order Pad'e-resummed WKB approximation and compared them to the eikonal prediction obtained from the photon-sphere angular frequency and the Lyapunov exponent. We find that increasing $\alpha$ systematically increases both $\mathrm{Re}(\omega)$ and $-\mathrm{Im}(\omega)$, giving faster and more strongly damped ringdown signals. The eikonal correspondence is correctly reproduced in the scalar and electromagnetic sectors, while larger deviations appear in the effective spin-2 channel due to the combined effect of the finite $\ell$ corrections and the phenomenological nature of the adopted Regge-Wheeler-like potential.

There is one particularly interesting result in the near extremal regime. Employing a Chebyshev collocation method together with a singular-value diagnostic, we find a branch of purely imaginary quasinormal frequencies associated with the near-horizon geometry. The damping rates of these modes show a weak dependence on the perturbation spin with relative differences below $1\%$. This suggests that the near-extremal spectrum is mostly controlled by the throat geometry and not by the spin-dependent structure of the effective potential. This behavior is very similar to the purely imaginary branch found for near-extremal Reissner-Nordstr\"om black holes and supports the interpretation that the near-horizon region gradually decouples from the normal photon-sphere branch as extremality is approached.

We also computed the exact greybody factors by numerical integration of the radial wave equation and compared them with analytical lower bounds. The PFDM contribution systematically suppresses transmission probabilities, increases backscattering and shifts the transition between reflection and transmission to higher frequencies. These modifications are directly manifest in the absorption cross sections and the Hawking emission spectra, where the dark matter halo acts as a reduction of the flux emitted in the Hawking radiation, even though it increases the Hawking temperature.

Finally, we studied the relativistic enhancement of the $\nu\bar{\nu}\rightarrow e^-e^+$ process. We have found that the PFDM environment significantly contributes to the energy deposition rate, while the Euler-Heisenberg correction induces only mild deviations, except for the near-extremal regime. This result suggests a possibly more important role of environmental effects than nonlinear electrodynamics in determining the energy budget available for high energy astrophysical phenomena powered by compact objects.

Overall, we find that the ringdown spectra, photon-sphere dynamics, greybody factors, Hawking radiation and neutrino energy deposition are all sensitive to the same underlying competition between compact non-linear-electrodynamic effects and extended dark matter environments. The EH+PFDM geometry is therefore a useful theoretical laboratory to investigate how environmental and strong-field corrections together affect black hole observables.

Several extensions are worth further investigation. To show that the effective spin-2 sector considered here is physically sound one would have to perform a fully coupled gravitational perturbation analysis. It would be interesting to study the effect of PFDM on tidal perturbations and echoes, investigate the rotating generalizations and look for possible observational implications for black hole spectroscopy in dark matter environments.


\begin{acknowledgments}

A. \"O.  would like to acknowledge networking support of the COST Action CA21106 - COSMIC WISPers in the Dark Universe: Theory, astrophysics and experiments (CosmicWISPers), the COST Action CA22113 - Fundamental challenges in theoretical physics (THEORY-CHALLENGES), the COST Action CA21136 - Addressing observational tensions in cosmology with systematics and fundamental physics (CosmoVerse), the COST Action CA23130 - Bridging high and low energies in search of quantum gravity (BridgeQG), and the COST Action CA23115 - Relativistic Quantum Information (RQI) funded by COST (European Cooperation in Science and Technology). Y. V. acknowledges the financial support of DIDULS/ULS, through the project No PR25538511.

\end{acknowledgments}

\end{document}